\documentclass[aps,prd,reprint,floatfix,superscriptaddress]{revtex4-2} 

\usepackage{graphicx}
\usepackage{dcolumn}
\usepackage{bm}

\usepackage{xcolor}
\usepackage{mathrsfs}
\usepackage{amsmath,amssymb}
\usepackage[pdftex]{hyperref}
\hypersetup{
  hidelinks
}

\begin{document}

\title{Expansion operators in spherically symmetric loop quantum gravity}

\author{Xiaotian Fei}
\affiliation{School of Physics and Astronomy, Key Laboratory of Multiscale Spin Physics
(Ministry of Education), Beijing Normal University, Beijing 100875, China}
\author{Gaoping Long}
\email{201731140005@mail.bnu.edu.cn}
\affiliation{College of Physics $\&$ Optoelectronic Engineering, Jinan University, Guangzhou, 510632, Guangdong, China}
\author{Yongge Ma}
\email{mayg@bnu.edu.cn}
\affiliation{School of Physics and Astronomy, Key Laboratory of Multiscale Spin Physics
(Ministry of Education), Beijing Normal University, Beijing 100875, China}
\author{Cong Zhang}
\email{cong.zhang@bnu.edu.cn}
\affiliation{School of Physics and Astronomy, Key Laboratory of Multiscale Spin Physics
(Ministry of Education), Beijing Normal University, Beijing 100875, China}

\date{\today}

\begin{abstract}
The ingoing and outgoing null expansions associated to a spatial 2-sphere are quantized in the spherically symmetric model of loop quantum gravity. It is shown that the resulting expansion operators are self-adjoint in the kinematical Hilbert space with generalized eigenstates. It turns out that the outgoing and ingoing expansion operators have the same spectrum consisting of a continuous band and discrete points. In the subspaces with fixed quantum numbers on the edges of the graphs underlining the quantum states, the two operators still share the common continuous spectrum, while their discrete spectra can be different from each other. These results provide new insights on the avoidance of the singularities in classical general relativity and the establishment of certain notion of quantum horizons.
\end{abstract}

\maketitle

\section{Introduction}

As a nonperturbative and background-independent approach to quantize gravity, loop quantum gravity (LQG) has achieved remarkable achievements, ranging from a mathematically precise kinematical framework to discrete spectra of geometric observables \cite{ashtekar2021short,rovelli2008loop,han2007fundamental,ashtekar2017loop,thiemann2008modern,ma2010new,bianchi2009length,dittrich2009spectra}. To study the issue of black hole entropy, the classical notion of isolated-horizon is employed to first reduce quasilocal degrees of freedom of the horizon, and then match the corresponding quantum theory on the horizon with the structure of LQG in the bulk \cite{engle2010black,ashtekar2004isolated,ashtekar1997quantum}. The horizon entropy obtained in this approach is proportional to the area of the horizon in leading order\cite{perez2017black,rovelli1996black,meissner2004black,agullo2010detailed,diaz2012isolated}. This picture can also be extended to dynamical horizons and related to the path-integral formulation of spinfoams \cite{pranzetti2012radiation,han2021spinfoam,ashtekar2020black}. In the dynamical procedure of black holes, such as the gravitational collapse, the mergers of black holes, and the evaporation of a black hole, the notions of apparent and trapping horizons, associated to a 2-sphere in a spatial manifold and its trajectory in the spacetime respectively, become very useful\cite{wald2010general,andersson2005local}. It is natural to expect certain kind of the notion of apparent horizon in quantum gravity theory like LQG. 

Symmetry-reduced models of LQG serve as simplified scenarios to test the structures and techniques in the full theory. The spherically symmetric midisuperspace model is particularly interesting since it still retains field-theoretic degrees of freedom, in contrast to the homogeneous minisuperspace models. To study the Schwarzschild black holes, the spherically symmetric reduction can be employed to treat the interior and the exterior within a single canonical framework, such that the gauge group of the connection formulation of general relativity (GR) is reduced from $\mathrm{SU}(2)$ to $\mathrm{U}(1)$ and the spatial diffeomorphisms are reduced to one-dimensional transformations while keeping the nontrivial constraint algebra. The resulting spherically symmetric theory can be polymer-like quantized by applying the method of LQG \cite{bojowald2006spherically,chiou2012loop,gambini2014quantum,gambini2020spherically,kelly2020effective}. Inspired by the spherically symmetric model of LQG, some effective models have been proposed to study the quantum properties of black holes. By employing certain polymer-modified Hamiltonian constraint as well as effective line element, the expansions of null geodesics in the effective spacetime can be calculated to locate apparent horizons. The quantum modifications to the horizons may be reflected by horizon shifts or multiplicities, regular cores, and mass-dependent corrections \cite{kelly2020effective,bojowald2018effective,bodendorfer2021mass,lewandowski2023quantum}. However, such analyses depend on the technical choices in the effective models, and the resulting notions are only valid at semiclassical level. Some fully quantum treatment and notion are desirable. In this paper, we consider the quantization of the expansions associated to a 2-sphere in the spherically symmetric model of LQG. The properties of the resulting expansion operators will be studied. Our purpose is to set up a starting point towards the quantum notion of horizons.

The paper is organized as follows. In section~\ref{sec:classical} the spherically symmetric model of classical connection formalism of GR and the definition of the ingoing and out going null expansions associated to a 2-sphere will be briefly reviewed. In section~\ref{sec:quantum} the spherically symmetric model of LQG will be introduced, and the null expansions will be quantized as well-defined operators in the kinematical Hilbert space. It turns out that the resulting expansion operators are self-adjoint. Section~\ref{sec:spectrum} will be devoted to study the properties of the expansion operators including their spectral analysis by both analytic and numerical methods. The results will be summarized with concluding remarks in the final section.

\section{ null expansions in spherically symmetric model  }\label{sec:classical}

Consider a spherically symmetric spacetime $(\mathscr{M},g_{ab})$ foliated by spatial slices $\Sigma_t\simeq I\times S^2$, with radial coordinate $x\in I$ and angle coordinates $(\theta,\phi)$ on the two-sphere $S^2$. All fields depend only on $x$ (and $t$). Recall that in the connection dynamics of GR, the canonical pairs consist of the $SU(2)$ connection $A^{i}{}_{a}$ and the densitized triad $\tilde{E}^{a}{}_{i}$ on the spatial manifold $\Sigma_t$, such that $A^i{}_{a}=\Gamma^i{}_{a}+\gamma K^i{}_{a},$ where $\Gamma^i{}_a$ is the spin connection compatible with the triad, $K^i{}_a$ is the $\mathfrak{su}(2)$-valued extrinsic-curvature one-form, and $\gamma$ is the Barbero--Immirzi parameter \cite{ashtekar1997quantum,thiemann2008modern}, and $\tilde{E}^a{}_{i}=\sqrt{q}\,e^{a}{}_{i}$, where $e^{a}{}_{i}$ is the triad, and $q$ is the determinant of the spatial metric on $\Sigma_t$. In the spherically symmetric model, the canonical pair is reduced to two pairs of fields $(E^x,K_x)$ and $(E^\varphi,K_\varphi)$ on $I$, where $E^\varphi$ and $K_x$ are scalar densities with weight one while $E^x$ and $K_\varphi$ are scalars\cite{chiou2012loop,bojowald2006spherically}. Thus the densitized triad reduces to two independent functions $E^x$ and $E^\varphi$ by choosing a symmetry-adapted triad basis aligned with the radial and angular directions, so that its non-vanishing components read
\[
\tilde E^x{}_3 = E^x \sin\theta,\qquad 
\tilde E^\theta{}_2 = E^\varphi \sin\theta,\qquad 
\tilde E^\phi{}_1 = E^\varphi .
\]
Hence the components of the densitized triad are related to the spatial metric $q_{ab}$ by
\begin{equation}
q_{xx}=\frac{(E^\varphi)^2}{|E^x|},\qquad
q_{\theta\theta}=|E^x|,\qquad q_{\phi\phi}=|E^x|\sin^2\theta.
\label{eq:q-metric}
\end{equation}
The non-vanishing extrinsic curvature components read 
\begin{equation}
\label{eq:K-cov}
\begin{aligned}
K_{xx} &= \frac{|E^\varphi|}{\sqrt{|E^x|}}\,K_x, \\
K_{\theta\theta} &= \operatorname{sgn}(E^\varphi)\sqrt{|E^x|}\,K_\varphi, \\
K_{\phi\phi} &= \operatorname{sgn}(E^\varphi)\sqrt{|E^x|}\,\sin^2\theta\,K_\varphi .
\end{aligned}
\end{equation}
Here $\operatorname{sgn}(\cdot)$ denotes the sign function with $\operatorname{sgn}(0)=0$. The non-trivial Poisson brackets among the reduced canonical variables are given by
\begin{equation}
\label{eq:PB-diag}
\begin{aligned}
\{K_x(x),E^x(x')\} &= 2G\gamma \,\delta(x-x'), \\
\{K_\varphi(x),E^\varphi(x')\} &= G\gamma \,\delta(x-x') .
\end{aligned}
\end{equation}
where $G$ is Newtonian gravitational constant, and we set the speed of light $c=1$.

With the intrinsic metric and extrinsic curvature of $\Sigma_t$ at hand, we can express the null expansions of two-spheres entirely in terms of the reduced canonical variables. For a smooth, oriented, closed $2$-sphere $S\subset\Sigma_t$ with outward unit normal $S^a$ tangent to $\Sigma_t$ and orthogonal to $S$, and with $n^a$ the future unit normal to $\Sigma_t$, the future-directed outward and inward null normals of $S$ are defined up to an irrelevant constant rescaling respectively by
\[
  k^a=n^a+S^a,\qquad
  l^a=n^a-S^a.
\]
Then, their expansions are given by \cite{wald2010general,evans1989frontiers}
\begin{align}
\Theta_{\rm out}(k) = D_a S^a + K_{ab}S^aS^b - K,
\label{eq:Theta-k}\\[0.25em]
\Theta_{\rm in}(l) = -\,D_a S^a + K_{ab}S^aS^b - K,
\label{eq:Theta-l}
\end{align}
where $D_a$ is the Levi--Civita connection of $q_{ab}$ and $K:=K^a{}_a$. In the spherical symmetric model, the outward unit normal of an 2-sphere $S$ defined by $x=\rm{constant}$ reads
\[
S^a=(q_{xx})^{-1/2}\Bigl(\frac{\partial}{\partial x}\Bigr)^a,
\]
and hence one has
\[
K_{ab}S^aS^b \;=\; K_{xx}\,(S^x)^2 \;=\; \frac{K_{xx}}{q_{xx}} \;=\; K^x{}_x \, .
\]
A direct calculation gives
\begin{equation}
D_a S^a
= \frac{\partial_x q_{\theta\theta}}{\sqrt{q_{xx}}\,q_{\theta\theta}}
= \frac{1}{\sqrt{|E^x|}\,|E^\varphi|}\,\partial_x\!\bigl(|E^x|\bigr) .
\label{eq:DivS}
\end{equation}
Combining Eqs.~\eqref{eq:q-metric}, \eqref{eq:PB-diag} and \eqref{eq:DivS}
with Eqs.~\eqref{eq:Theta-k} and \eqref{eq:Theta-l}, we obtain expressions of the expansions of $S$ in the spherically symmetric model as
\begin{align}
\Theta_{\rm out}
= \frac{1}{\sqrt{|E^x|}\,|E^\varphi|}\,\partial_x\!\bigl(|E^x|\bigr)
 - \frac{2\operatorname{sgn}(E^\varphi)}{\sqrt{|E^x|}}\,K_\varphi, \label{eq:Theta-out}\\[0.25em]
\Theta_{\rm in}
= -\,\frac{1}{\sqrt{|E^x|}\,|E^\varphi|}\,\partial_x\!\bigl(|E^x|\bigr)
 - \frac{2\operatorname{sgn}(E^\varphi)}{\sqrt{|E^x|}}\,K_\varphi. \label{eq:Theta-in}
\end{align}

Note that the line element of the spacetime metric in the coordinate system compatible with $S$ can be written as
\begin{equation}
\mathrm{d}s^2=-N^2\,\mathrm{d}t^2+q_{xx}\bigl(\mathrm{d}x+N_r\,\mathrm{d}t\bigr)^2
+q_{\theta\theta}\,\mathrm{d}\theta^2+q_{\phi\phi}\,\mathrm{d}\phi^2 ,
\label{eq:ADMline}
\end{equation}
with lapse $N$ and radial shift $N_r$. For the Schwarzschild slicing, in the exterior region $x>2GM$, one has $K_\varphi=0$, $E^x=x^2$ and $E^\varphi=\frac{x}{\sqrt{1-2GM/x}}$ (see, e.g., Refs.\cite{bojowald2006spherically,chiou2012loop}). Then we obtain the expansions of $S$ as
\[
\Theta_{\rm out}=\frac{2}{x}\sqrt{1-\frac{2GM}{x}},\qquad
\Theta_{\rm in}=-\,\frac{2}{x}\sqrt{1-\frac{2GM}{x}},
\]
thus that $\Theta_{\rm out}\to 0$ as $x\to(2GM)^+$. For the
Painlev\'e--Gullstrand slicing in which the horizon can be covered by the coordinate system, one has $E^x=x^2$, $E^\varphi=x$ and
$K_\varphi=\sqrt{2GM/x}$ \cite{bojowald2006spherically,chiou2012loop}. Then we obtain the expansions
\[
\Theta_{\rm out}=\frac{2}{x}\Bigl(1-\sqrt{\frac{2GM}{x}}\Bigr),\qquad
\Theta_{\rm in}=-\,\frac{2}{x}\Bigl(1+\sqrt{\frac{2GM}{x}}\Bigr),
\]
and hence $\Theta_{\rm out}=0$ at the horizon $x=2GM$.

\section{Quantum Theory}\label{sec:quantum}
\subsection{Kinematical structure}\label{subsec:kinematics}

As in the full theory of LQG, the loop quantization of the spherically symmetric model can be realized by representing the holonomy--flux algebra in a kinematic Hilbert space $\mathscr{H}_{\mathrm{kin}}$. Specifically, as the variable $K_x$ transforms as an $U(1)$-connection under residual gauge transformations in the procedure of symmetric reduction\cite{gambini2014quantum,zhang2019bouncing,kelly2020black}, the holonomy--flux algebra is generated by the $U(1)$-holonomies of $K_x$ along suitable paths, point holonomies of $K_\varphi$ at vertices, and the fluxes of $E^x$ and $E^\varphi$. The kinematical Hilbert space $\mathscr{H}_{\mathrm{kin}}$ is spanned by cylindrical functions supported on all possible finite graphs $g\subset I$, where the edges of a graph run along the radial line and the vertices mark the points that carry the information of the angular directions. 

A given graph $g$ consists of finite numbers of oriented, non-overlapping edges $\{e_j\}$ aligned with $I$ and finite numbers of vertices $\{v_j\}$. Once the orientation of $I$ is fixed, all edges inherit that orientation. We associate the radial connection $K_x$ to each edge and the angular scalar $K_\varphi$ to each vertex. Then, by following the loop quantization method \cite{thiemann1998kinematical,ashtekar2002polymer}, the kinematical Hilbert space $\mathscr{H}^{g}_{\mathrm{kin}}$ on graph $g$ is given by
\begin{equation}
\mathscr{H}^{g}_{\mathrm{kin}}
= \left(\,\bigotimes_{e_j\in g}\ \ell^2(\mathbb{Z})\,\right)\ \otimes\
  \left(\,\bigotimes_{v_j\in g}\ L^2\!\big(\mathbb{R}_{\mathrm{Bohr}},\,\mathrm{d}\mu_{\mathrm{Bohr}}\big)\,\right),
\label{eq:kin-Hg}
\end{equation}
where the Hilbert space $\ell^2(\mathbb{Z})$ associate to an edge consists of square-summable sequences on the set of integers, $\mathbb{Z}$, equipped with the counting measure, and the Hilbert space $L^2(\mathbb{R}_{\mathrm{Bohr}},\mathrm{d}\mu_{\mathrm{Bohr}})$ associate to a vertex consists of square-integrable almost-periodic functions on space $\mathbb{R}_{\mathrm{Bohr}}$ of the Bohr compactification  with respect to the normalized Haar measure $\mathrm{d}\mu_{\mathrm{Bohr}}$. 
A basis in $\mathscr{H}^{g}_{\mathrm{kin}}$ takes the form
\begin{equation}
\label{eq:kin-cyl}
\begin{aligned}
T_{g,\vec{k},\vec{\mu}}(K_x,K_{\varphi})
&= \prod_{e_j \in g} \exp\!\left(\frac{\mathrm{i}\,k_j}{2}\int_{e_j} K_x(x)\,\mathrm{d}x\right) \\
&\quad \times \prod_{v_j \in g} \exp\!\left(\mathrm{i}\,\mu_j\,K_{\varphi}(v_j)\right) .
\end{aligned}
\end{equation}
with edge labels $k_j\in\mathbb{Z}$ and vertex labels $\mu_j\in\mathbb{R}$, where $\vec{k}=(\dots,k_{j-1},k_j,k_{j+1},\dots)$, and $\vec{\mu}=(\dots,\mu_{j-1},\mu_j,\mu_{j+1},\dots)$.
The full kinematical Hilbert space is the orthogonal direct sum
\begin{equation}
\mathscr{H}_{\mathrm{kin}}=\bigoplus_{g}\ \mathscr{H}'^{g}_{\mathrm{kin}}\bigoplus\mathscr{C},
\label{eq:kin-H}
\end{equation}
with inner product
\begin{equation}
\langle g, \vec{k}, \vec{\mu}\,|\,g', \vec{k}', \vec{\mu}'\rangle
= \delta_{g,g'}\,\delta_{\vec{k},\vec{k}'}\,\delta_{\vec{\mu},\vec{\mu}'},
\label{eq:kin-ip}
\end{equation}
where $\mathscr{H}'^{g}_{\mathrm{kin}}$ is spanned by the cylindrical functions $T'_{g,\vec{k},\vec{\mu}}(K_x,K_{\varphi})$ with nontrivial $k_j$ on any edge and nontrivial $\mu_j$ on any vertex separating the two edges with the same label $k_j$, $\mathscr{C}$ denotes the constant function corresponding to a trivial graph, and the states $|g,\vec{k},\vec{\mu}\rangle$ correspond to the normalized version of $T'_{g,\vec{k},\vec{\mu}}(K_x,K_{\varphi})$. The basis$|\,g, \vec{k}, \vec{\mu}\rangle$, including the trivial graph, spans a dense domain that will also serve as the natural domain for the operators introduced below.

The operators corresponding to $E^x$ and $E^\varphi$ act on the basis $|g,\vec{k},\vec{\mu}\rangle$ diagonally. Since $E^x$ is a scalar on $I$, we can promote $E^x(p)$ at an arbitrary point $p$ directly to an operator $\widehat{E}^x(p)$. If $p$ lies in the interior of an edge $e_j$ of a quantum state, the action $\widehat{E}^x(p)$ on the state $\lvert g,\vec k,\vec\mu\rangle$ simply reads off the label $k_j$, so that $\lvert g,\vec k,\vec\mu\rangle$ is an eigenstate of $\hat E^x(p)$ with eigenvalue propotional to $k_j$\cite{Zhang:2021xoa}. If $p$ coincides with a vertex connecting edges $e_j$ and $e_{j+1}$ of a quantum state, an ambiguity arises as to whether the eigenvalue should be $k_j$ or $k_{j+1}$. Several prescriptions for the regularization of the action exist in the literature, and in general they lead to inequivalent quantum operators\cite{chiou2012loop,gambini2020spherically}. Here we adopt the symmetric prescription, in which $\hat E^x(p)$ takes the average of the two labels $k_j$ and $k_{j=1}$. This choice is more naturally aligned with the corresponding flux regularization in the full theory of LQG\cite{thiemann2008modern,ashtekar2005100}. More precisely, $\hat E^x(p)$ acts on a basis state whose graph contains $p$ as
\begin{equation}
\label{eq:kin-Ex}
\widehat{E}^{x}(p)\,|g,\vec{k},\vec{\mu}\rangle
= \gamma\,l_p^2\, \bar s_j(p)\,|g,\vec{k},\vec{\mu}\rangle,
\end{equation}
with
\[
\bar s_j(p) =
\begin{cases}
{\bar k_j}:=\dfrac{k_j+k_{j+1}}{2}, & p=v_j=t(e_j)=s(e_{j+1}),\\[0.6em]
k_j, & p\in e_j\setminus\{s(e_j),t(e_j)\},
\end{cases}
\]
where $t(e_j)$ is the target vertex of $e_j$, $s(e_j)$ is its source vertex, and $l_p^2=G\hbar$ is the Planck area. As $E^\varphi$ is a density on $I$, for an {open} interval $\mathcal{I}\subset I$, we define the smeared quantity $E^{\varphi}(\mathcal I):=\int_{\mathcal I} E^{\varphi}(x)\,\mathrm{d}x$. Its corresponding operators act on the basis as
\begin{equation}
\widehat{E}^{\varphi}(\mathcal{I})\,|g,\vec{k},\vec{\mu}\rangle
= \gamma \,l_p^2 \sum_{v_j\in g\cap \mathcal{I}} \mu_j\,|g,\vec{k},\vec{\mu}\rangle.
\label{eq:kin-Ephi-I}
\end{equation}
The holonomy of $K_x$ along $e_j$  takes the form
\begin{equation}
N^x_{\pm}(e_j):=\exp\!\Big[\pm\frac{\mathrm{i}}{2}\int_{e_j}  K_x\,\mathrm{d}x\Big].
\label{eq:kin-Nx}
\end{equation}
Its corresponding operator acts on quantum states by a multiplication. In particular, if the edge $e_j$ belong to a basis vector, the action reads 
\begin{equation}
\label{eq:kin-Nx-vertex}
\begin{aligned}
&\widehat{N}^x_{\pm}(e_j)\,|g,\vec{k},\vec{\mu}\rangle
= |g,\vec{k}'_{\pm},\vec{\mu}\rangle, \\
&\vec{k}'_{\pm} = (\dots,k_{j-1},k_j\!\pm\!1,k_{j+1},\dots).
\end{aligned}
\end{equation}
For latter convenience, the point holonomies of $K_\varphi$ at $p\in I$ are defined as
\begin{equation}
N^{\varphi}_{\pm n\rho}(p):=\exp\!\left\{\mathrm{i}\,(\pm n\rho)\,K_{\varphi}(p)\right\},\qquad n\in\mathbb{N}.
\label{eq:kin-Nphi}
\end{equation}
The corresponding operators act on quantum states by multiplications. In particular, if the point $p$ belongs to the graph underlying a basis state,
the action reads
\begin{equation}
\label{eq:9}
\begin{aligned}
&\widehat{N}^{\varphi}_{\pm n\rho}(p)\,|g,\vec k,\vec{\mu}\rangle
\\
&\qquad=\begin{cases}
|g,\vec{k},\vec{\mu}'_{\pm n\rho}\rangle,
& \text{if } p=v_j\in g,\\[0.4em]
|g',\vec{k}_p,\vec{\mu}'_{0\pm n\rho}\rangle,
& \text{if } p\in e_j\setminus\{s(e_j),t(e_j)\}.
\end{cases}
\end{aligned}
\end{equation}
where $g'$ is the refined graph obtained by inserting $p$ as a new vertex on $e_j$,
$\vec{k}_p:=(\dots,k_{j-1},k_j,k_j,k_{j+1},\dots)$, and
\[
\begin{aligned}
\vec{\mu}'_{\pm n\rho}
&:= (\dots,\mu_{j-1},\mu_j\!\pm\! n\rho,\mu_{j+1},\dots),\\
\vec{\mu}'_{0\pm n\rho}
&:= (\dots,\mu_{j-1},\mu_j,\mu_p=\pm n\rho,\mu_{j+1},\dots).
\end{aligned}
\]
Note that the point holonomies $N^{\varphi}_{\pm n\rho}(p)$ should be constructed to represent the holonomies of the connection along the angular directions in the full theory. As argued in Refs.\cite{zhang2022loop,kelly2020effective,lewandowski2023quantum}, the so-called  $\bar\mu$-scheme is preferred in the construction of the point holonomies. In this scheme the \emph{physical} length of a segment along an angular direction is fixed by the area gap of the full theory as $\delta=2\sqrt{2\sqrt{3}\pi\gamma}\,l_p$. Hence, adapting to the $\bar{\mu}$-scheme,
the \emph{coordinate} step in the construction of the point holonomies can be given by
\begin{equation}
\tilde{\rho}=\frac{\delta}{\sqrt{|E^x|}}\, .
\label{eq:7}
\end{equation}
With this choice, Eq.~\eqref{eq:9} can be rewritten as
\begin{equation}
\label{eq:Nphi}
\begin{aligned}
&\widehat{N}^{\varphi}_{\pm n\tilde{\rho}}(p)\,|g,\vec{k},\vec{\mu}\rangle
\\
&\qquad=
\begin{cases}
\begin{aligned}
|g,\vec{k},&\vec{\mu}'_{\pm n\rho_j}\rangle,
\text{if } p=v_j\in g,\\
\end{aligned}
\\[0.4em]
|g',\vec{k}_p,\vec{\mu}'_{0\pm n\rho_p}\rangle,
\ \text{if } p\in e_j\setminus\{s(e_j),t(e_j)\},
\end{cases}
\end{aligned}
\end{equation}
where $\rho_j$ takes the eigenvalue of $\tilde{\rho}$ at a true vertex $v_j$ as
\begin{equation}
\label{eq:rho-eig}
\rho_j
= 2\,\sqrt{2\sqrt{3}\pi}\;
\operatorname{sgn}({\bar k_j})\,
\Bigl(|{\bar k_j+1}|^{1/2}-|{\bar k_j-1}|^{1/2}\Bigr),
\end{equation}
while for a point $p$ in the interior of an edge $e_j$ one has
\begin{equation}
\label{eq:rho-eig-p}
\rho_p
= 2\,\sqrt{2\sqrt{3}\pi}\;
\operatorname{sgn}(k_j)\,
\Bigl(|k_j+1|^{1/2}-|k_j-1|^{1/2}\Bigr).
\end{equation}
Since $\widehat E^{x}(p)$ and $\widehat E^{\varphi}(\mathcal I)$ are self-adjoint and diagonal on $|g,\vec k,\vec\mu\rangle$, the corresponding sign operators can be defined by their action on the eigenbasis as
\begin{equation}
\label{eq:sgn-action-basis}
\begin{aligned}
\widehat{\operatorname{sgn}(E^{x})}(p)\,|g,\vec{k},\vec{\mu}\rangle
&:= \operatorname{sgn}\!\left(\gamma\, l_p^2\, {\bar s_j(p)}\right)\,
|g,\vec{k},\vec{\mu}\rangle, \\[0.4em]
\widehat{\operatorname{sgn}(E^{\varphi})}(\mathcal{I})\,|g,\vec{k},\vec{\mu}\rangle
&:= \operatorname{sgn}\!\left(\gamma\, l_p^2
\sum_{v_j\in g\cap \mathcal{I}} \mu_j \right)\,
|g,\vec{k},\vec{\mu}\rangle .
\end{aligned}
\end{equation}
These sign operators are bounded and self-adjoint. They coincide with the operators
obtained by applying the spectral theorem to $\widehat E^{x}(p)$ and
$\widehat E^{\varphi}(\mathcal{I})$.

\subsection{Expansion operators}\label{subsec:expansion-operator}

We now promote the classical null expansions to operators in
$\mathscr{H}_{\rm kin}$. As shown in Eqs.~\eqref{eq:Theta-out} and \eqref{eq:Theta-in},
the variable $K_\varphi$ explicitly appears in the expressions of the expansions while it does not correspond to a well-defined operator in
$\mathscr{H}_{\rm kin}$. Thus, we adopt the $\bar{\mu}$ scheme to regularize $K_\varphi$ as
\begin{equation}
\label{eq:Pk}
P(K_\varphi)\equiv \frac{\sin(\tilde{\rho} K_\varphi)}{\tilde{\rho}}.
\end{equation}
Moreover, the other terms in Eqs.~\eqref{eq:Theta-out} and \eqref{eq:Theta-in} can be written in the forms suitable for quantization by the following identities:
\begin{align}
\frac{1}{\sqrt{|E^x|}}(p)
&=\frac{2\,\operatorname{sgn}(E^x)}{\mathrm{i}\gamma G}
\left\{\sqrt{|E^x|},\,{N^x_{+}(\mathcal{I}^\epsilon)}\right\}
{N^x_{-}(\mathcal{I}^\epsilon)},
\label{eq:Ex}
\\
\frac{|E^x|'}{|E^\varphi|}(p)
&= \frac{1}{G^2\gamma^2\,\tilde{\rho}^2}
\lim_{\epsilon\to 0}\!
\Bigl[
2\bigl\{\sqrt{|E^\varphi|(\mathcal{I}^\epsilon)},\,N^\varphi_{+\tilde{\rho}}\bigr\}
\,N^\varphi_{-\tilde{\rho}}
\Bigr]^{2}
\nonumber\\
&\quad \times
\int_{\mathcal{I}^\epsilon}\partial_x |E^x|\,\mathrm{d}x,
\label{eq:Ex'}
\end{align}
where $\mathcal{I}^\epsilon$ denotes {an open} coordinate interval of length $\epsilon$ containing the point $p$, {and $N^x_{\pm}(\mathcal{I}^\epsilon)$ denotes the holonomies of $K_x$ along the interval $\mathcal{I}^\epsilon$}. Now the expressions of Eqs.~\eqref{eq:Pk} and \eqref{eq:Ex} have got clear quantum analogs. To quantize Eq.~\eqref{eq:Ex'}, we first take the integral and then quantize its quantities. It turns out that the regularized operator has nontrivial action only on the edges of the graph underlying a cylindrical function in $\mathscr{H}_{\rm kin}$, and the limit can be taken trivially. Thus the null expansions can be consistently promoted to operators in $\mathscr{H}_{\rm kin}$ as
\begin{align}
\widehat{\Theta}_{\text{out}}
&:= \widehat{\left[\frac{|E^x|'}{|E^\varphi|}\right]}\,
   \widehat{\left[\frac{1}{\sqrt{|E^x|}}\right]}
 \;-\; 2\,\widehat{\left[\frac{1}{\sqrt{|E^x|}}\right]}\,\widehat{\mathcal{P}}_{\varphi},
\label{eq:ThetaOut-op-def}\\
\widehat{\Theta}_{\text{in}}
&:=  -\,\widehat{\left[\frac{|E^x|'}{|E^\varphi|}\right]}\,
     \widehat{\left[\frac{1}{\sqrt{|E^x|}}\right]}
 \;-\; 2\,\widehat{\left[\frac{1}{\sqrt{|E^x|}}\right]}\,\widehat{\mathcal{P}}_{\varphi},
\label{eq:ThetaIn-op-def}
\end{align}
where the operators corresponding to Eqs.~\eqref{eq:Ex} and \eqref{eq:Ex'} act on the basis states respectively as
\begin{equation}
\begin{aligned}
\widehat{\left[\frac{1}{\sqrt{|E^x|}}\right]}|g,\vec{k},\vec{\mu}\rangle
&=\frac{\operatorname{sgn}({\bar s_j})}{\sqrt{\gamma}\,l_p}
\Bigl(|{\bar s_j+1}|^{\tfrac12}-|{\bar s_j-1}|^{\tfrac12}\Bigr)
\\
&\quad \times |g,\vec{k},\vec{\mu}\rangle,\\[0.3em]
\widehat{\left[\frac{|E^x|'}{|E^\varphi|}\right]}(v_j)\,|g,\vec{k},\vec{\mu}\rangle
&= \frac{1}{\rho_j^{2}}\,
\Delta\Bigl(|\mu_j+\rho_j|^{\tfrac12}-|\mu_j-\rho_j|^{\tfrac12}\Bigr)^{2}
\\
&\quad \times |g,\vec{k},\vec{\mu}\rangle .
\end{aligned}
\label{eq:poly-Kphi}
\end{equation}
with $\Delta:=|k_{j+1}|-|k_j|$ and $\rho_j$ given by Eq.~\eqref{eq:rho-eig}, and the symmetric operator
\begin{equation}
\widehat{\mathcal{P}}_{\varphi}
:=\frac12\Big(
\widehat{\operatorname{sgn}(E^\varphi)}\,\widehat{P(K_\varphi)}
+\widehat{P(K_\varphi)}\,\widehat{\operatorname{sgn}(E^\varphi)}
\Big),
\label{eq:Pphi-def}
\end{equation}
is introduced with
\begin{equation}
\widehat{P(K_\varphi)}:=\frac{\hat N^\varphi_{+\tilde\rho}-\hat N^\varphi_{-\tilde\rho}}{2\,\mathrm{i}\,\tilde\rho}.
\end{equation}

Since the operators $\widehat{\Theta}_{\text{in}}$ and $\widehat{\Theta}_{\text{out}}$ differ only by the sign in front of the factor $\widehat{\big[|E^x|'/|E^\varphi|\big]}$, the analysis of their operator properties proceed in parallel. Moreover, classically an apparent horizon is characterized by the vanishing of its outgoing expansion in the spherically symmetric geometry. Hence in what follows we will focus on $\widehat{\Theta}_{\text{out}}$, while the properties of $\widehat{\Theta}_{\rm{in}}$ will be mentioned as remarks. According to the definition \eqref{eq:ThetaOut-op-def}, at a point $v_j$ on the graph where the adjacent edge labels have a nontrivial jump $\Delta\neq 0$, the operator $\widehat{\Theta}_{\text{out}}$ acts on the basis states as
\begin{equation}
\label{eq:theta1}
\begin{aligned}
&\widehat{\Theta}_{\rm out}(v_j)\,|g,\vec{k},\vec{\mu}\rangle
\\
&= \frac{\operatorname{sgn}({\bar s_j})\Bigl(|\mu_j+\rho_j|^{\tfrac12}-|\mu_j-\rho_j|^{\tfrac12}\Bigr)^{2}}
{8\sqrt{3\gamma}\pi\,l_p\,
\Bigl(|{\bar s_j+1}|^{\tfrac12}-|{\bar s_j-1}|^{\tfrac12}\Bigr)}
\,\Delta|g,\vec{k},\vec{\mu}\rangle\\
&\qquad-\alpha
[\operatorname{sgn}(\mu_j)+\operatorname{sgn}(\mu_j+\rho_j)]\,|g,\vec{k},\vec{\mu}'_{+\rho_j}\rangle \\
&\qquad+\alpha[\operatorname{sgn}(\mu_j)+\operatorname{sgn}(\mu_j-\rho_j)]\,|g,\vec{k},\vec{\mu}'_{-\rho_j}\rangle.
\end{aligned}
\end{equation}
where $\alpha=\frac{1}{4\sqrt{2\sqrt{3}\pi\gamma}\,\mathrm{i}\,l_p}$. Whenever $\widehat{\Theta}_{\text{out}}$ acts at a point $p$ with $\Delta=0$, the diagonal term proportional to $\Delta$ is absent and the action reduces to a pure shift of the corresponding $\mu$-label. For example, this occurs if $p$ lies in the interior of an edge $e_j$. In this case, the action transforms label $\mu_p$ of $p$ as
\begin{equation}
\label{eq:theta2}
\begin{aligned}
&\widehat{\Theta}_{\rm out}(p)\,|g',\vec{k},\vec{\mu}_p\rangle
\\
&= -\alpha
[\operatorname{sgn}(\mu_p)+\operatorname{sgn}(\mu_p+\rho_p)]\,|g',\vec{k},\vec{\mu}'_{p+\rho_p}\rangle 
\\
&+\alpha[\operatorname{sgn}(\mu_p)+
\operatorname{sgn}(\mu_p-\rho_p)]|g',\vec{k},\vec{\mu}'_{p-\rho_p}\rangle.
\end{aligned}
\end{equation}
where $g'$ is either the graph obtained by inserting $p$ as a new vertex on $e_j$, or $g'=g$ if $p$ is already a vertex of $g$. Note that the action of the operator $\widehat{\Theta}_{\rm{in}}$ on the basis states takes the same form as Eqs.~\eqref{eq:theta1} and \eqref{eq:theta2}, but there would be a sign difference in the first term in Eq.~\eqref{eq:theta1}. {However, for any state $\lvert g',\vec k,\vec \mu\rangle$ with fixed $\vec k$, one can always choose another state with label assignment $\vec k'=(\dots,k_{j-1},k_{j+1},k_j,\dots)$ such that $\widehat\Theta_{\rm in}(v_j)\,\lvert g',\vec k',\vec \mu\rangle$ takes exactly the same coefficient structure as $\widehat\Theta_{\rm out}(v_j)\,\lvert g',\vec k,\vec \mu\rangle$. This mirrors the classical picture: on a fixed background (corresponding here to fixing $g$ and $\vec k$), the ingoing and outgoing expansions at the same point generally differ. However, once one allows all backgrounds, the ingoing expansion in one background can be identified with the outgoing expansion in another, related by an orientation flip.}

To study the properties of the operator $\widehat\Theta_{\rm out}(v)$ at a point $v$, without lose of generality, we ask all the graphs underlining the cylindrical functions to be adapted to $v$ in the sense that $v$ is always a vertex of the graphs.
This implies that $v$ could be a virtual vertex with the quantum number $\mu_v=0$. Thus graphs are denoted by $g_v$. Let $\mathscr{H}'^{g_v}_{\rm kin}$ be the Hilbert space over $g_v$ spanned by the cylindrical functions $T'_{g_v,\vec{k},\vec{\mu}}$ with nontrivial $k_j$ on any edge and nontrivial $\mu_j$ on any vertex except for $v$. 
Then Eqs.~\eqref{eq:theta1} and \eqref{eq:theta2} ensure that $ \widehat{\Theta}_{\text{out}}(v)$ is an operator in each $\mathscr{H}^{g_v}_{\rm kin}$.
Moreover, the full kinematical Hilbert space admits the following orthogonal decomposition:
\begin{equation}\label{eq:rearranged-sum}
\mathscr H_{\rm kin}
=\bigoplus_{g_v}\,\mathscr H^{g_v}_{\rm kin}\ \oplus\ \mathscr C .
\end{equation}
We denote the restriction of the operate $\widehat\Theta_{\rm out}(v)$ in $H^{g_v}_{\rm kin}$ as $\widehat\Theta_{\rm out}^{g_v}:=\widehat\Theta_{\rm out}(v)\!\restriction_{\mathscr H^{g_v}_{\rm kin}}$.
Eqs.~\eqref{eq:theta1} and \eqref{eq:theta2} also imply that in a given $\mathscr H^{g_v}_{\rm kin}$, the operator $\widehat\Theta_{\rm out}^{g_v}$ preserves the edge-labels $\vec k$ of the cylindrical functions. Let
$\mathscr H^{g_v}_{\vec k}$ be the closed subspace in $\mathscr H^{g_v}_{\rm kin}$ spanned by
the basis states $|g,\vec k,\vec\mu\rangle$ with $\vec k$ fixed. Then the Hilbert space $\mathscr H^{g_v}_{\rm kin}$ admits the following orthogonal decomposition:
\[
\mathscr H^{g_v}_{\rm kin}=\bigoplus_{\vec k}\,\mathscr H^{g_v}_{\vec k}.
\]
Accordingly, the operator $\widehat\Theta_{\rm out}^{g_v}$ can also be decomposed as
\begin{equation}\label{eq:direct-sum-s}
\widehat\Theta_{\rm out}^{g_v}
=\bigoplus_{\vec k}\,\widehat\Theta_{\rm out}^{(g_v,\vec k)} 
\end{equation}
with
\begin{equation}\label{eq:theta-g-k}
\qquad
\widehat\Theta_{\rm out}^{(g_v,\vec k)}
:=\widehat\Theta_{\rm {out}}^{g_v}\!\restriction_{\mathscr H^{g_v}_{\vec k}} .
\end{equation}
As shown in Appendix~\ref{app:appa}, the operator $\widehat\Theta_{\rm out}^{(g_v,\vec k)}$ is symmetric and satisfies the bound
\begin{equation}\label{eq:bounded-s}
\|\widehat{\Theta}^{(g_v,\vec k)}_{\rm out}\varphi\|
\le \frac{1}{\sqrt{\gamma}\,l_p}\,
\sqrt{\frac{6\,\Delta^2+3}{4\sqrt{3}\,\pi}}\;\|\varphi\|,
\qquad \forall\,\varphi\in\mathscr H^{g_v}_{\vec k}.
\end{equation}
Therefore, it is self-adjoint on $\mathscr H^{g_v}_{\vec k}$. Thus by the direct-sum theorem for self-adjoint operators\cite{reed1978iv}, $\widehat\Theta_{\rm out}^{g_v}$ is self-adjoint on its natural domain in $\mathscr H^{g_v}_{\rm kin}$. Moreover, since the operator $\widehat\Theta_{\rm out}(v)$ vanishes the constant functions, by combining the decomposition \eqref{eq:rearranged-sum} with the same direct-sum theorem,
$\widehat\Theta_{\rm out}(v)$ is self-adjoint in $\mathscr H_{\rm kin}$ with the natural domain
\begin{equation}\label{eq:domain-full-proj}
\mathcal D(\widehat\Theta_{\rm out}(v))
=
\left\{
\Psi\in\mathscr H_{\rm kin}\ \Big|\ 
\bigl\|\widehat\Theta_{\rm out}^{g_v}(v)\,\Psi\bigr\|^2<\infty
\right\}
\end{equation}
which is dense in $\mathscr H_{\rm kin}$. The above discussions are applicable in parallel to the operator $\widehat\Theta_{\rm in}$. Its restriction to $\mathscr H^{g_v}_{\vec k}$ is also symmetric and satisfies the same bound \eqref{eq:bounded-s}, and hence $\widehat\Theta_{\rm in}$ is also self-adjoint in $\mathscr H_{\rm kin}$.

\section{Spectra of the Expansion operators}\label{sec:spectrum}
Based on Eqs.~\eqref{eq:theta1} and \eqref{eq:theta2}, we now derive the corresponding (generalized) eigen equations of $\widehat{\Theta}_{\text{out}}$ in the $\mu$-representation. Without loss of generality, we express the eigen equations in terms of the generalized states belonging to the dual space of a dense set in $\mathscr{H}_{\mathrm{kin}}$ and denoted by $(\cdot|$. Note that any eigenstate of $\widehat{\Theta}_{\text{out}}$ falls into one of the following two cases, depending on whether $\Delta\neq0$ or $\Delta=0$ at the acting point. At a point $v_j$ with $\Delta\neq 0$, as the action of $\widehat{\Theta}_{\text{out}}$ changes only the $\mu$-label at $v_j$, a generalized eigenstate obeys
\begin{equation}\label{eigenequation111}
\left(g, \omega_{v_j},\vec{\mu}_{\setminus v_j},\vec{k}\right|\,\widehat{\Theta}^{\dagger}_{\text{out}}(v_j)
= \omega_{v_j}\,\left( g,\omega_{v_j},\vec{\mu}_{\setminus v_j},\vec{k}\right|,
\end{equation}
with $\vec{\mu}_{\setminus v_j}=(\dots,\mu_{j-1},\mu_{j+1},\dots)$. The eigenstate can be expanded in the spin-network basis as
\begin{equation}
\left( g,\omega_{v_j},\vec{\mu}_{\setminus v_j},\vec{k}\right|
= \sum_{\mu_j}\,\langle g,\vec{k},\vec{\mu}|\,\phi_{\omega_{v_j},\vec{k}}(\vec{\mu}).
\label{eq:8}
\end{equation}
Then, by combining Eqs.~\eqref{eq:theta1} and \eqref{eigenequation111}, we get a finite-difference equation for the coefficients $\phi_{\omega_{v_j},\vec{k}}(\vec{\mu})$ as
\begin{widetext}
\begin{equation}
\label{eq:eigenequationv}
\begin{aligned}
\Biggl(
\frac{\operatorname{sgn}({\bar s_j})\Bigl(|\mu_j+\rho_j|^{\tfrac12}-|\mu_j-\rho_j|^{\tfrac12}\Bigr)^{2}}
{8\sqrt{3\gamma}\pi\,l_p\,
\Bigl(|{\bar s_j+1}|^{\tfrac12}-|{\bar s_j-1}|^{\tfrac12}\Bigr)}
\,\Delta
-\omega_{v_j}
\Biggr)\,
\phi_{\omega_{v_j},\vec{k}}(\mu_j)&
\\
-\alpha\Bigl(\,[\operatorname{sgn}(\mu_j)+\operatorname{sgn}(\mu_j+\rho_j)]\,
\phi_{\omega_{v_j},\vec{k}}(\mu_j+\rho_j)
&-\,[\operatorname{sgn}(\mu_j)+\operatorname{sgn}(\mu_j-\rho_j)]\,
\phi_{\omega_{v_j},\vec{k}}(\mu_j-\rho_j)\Bigr)
=0 .
\end{aligned}
\end{equation}
\end{widetext}
At a point $p$ with $\Delta=0$, the corresponding eigenstate of $\widehat{\Theta}_{\text{out}}(p)$ can be expressed as $\sum_{\mu_p}\langle g',\vec{k},\vec{\mu}|\,\phi_{\omega_{p},\vec{k}}(\vec{\mu})$, and the coefficients satisfy the difference equation
\begin{equation}
\label{eq:5}
\begin{aligned}
&\omega_{p}\,\phi_{\omega_{p},\vec{k}}(\mu_p)
+\alpha\,[\operatorname{sgn}(\mu_p)+\operatorname{sgn}(\mu_p+\rho_p)]\,
\phi_{\omega_{p},\vec{k}}(\mu_p+\rho_p)
\\
&-\alpha\,[\operatorname{sgn}(\mu_p)+\operatorname{sgn}(\mu_p-\rho_p)]\,
\phi_{\omega_{p},\vec{k}}(\mu_p-\rho_p)
=0 .
\end{aligned}
\end{equation}
It should be noted that, since $\widehat\Theta_{\rm out}$ is self-adjoint, its spectrum belongs to the field of real numbers.

\subsection{The spectra in the sector of $\Delta=0$}\label{sec:spectrum1}
Since the eigenstates of $\widehat\Theta_{\rm out}$ are naturally divided into two sectors of $\Delta=0$ and $\Delta\neq0$. It is convenient to discuss its spectrum in the two sectors separately. In the sector of $\Delta=0$, the eigenequation \eqref{eq:5} of $\widehat\Theta_{\rm out}$ is a second-order difference equation on the $\mu$-lattice. Since the step of the $\mu$-lattice is $\rho_p$, there always exists a point $\mu_p=\mathring\mu_p$ satisfying $0\leq \mathring \mu_p<\rho_p$ on such a lattice.
The solution to the eigenequation Eq.~\eqref{eq:5} is completely determined once the initial values  $\phi_{\omega_{p},\vec{k}}(\mathring \mu_p)$ and $\phi_{\omega_{p},\vec{k}}(\mathring \mu_p-\rho_p)$ are specified. This becomes explicit by simplifying Eq.~\eqref{eq:5} as follows:
\begin{itemize}
    \item[(i)] For the case of $\mathring \mu_p\neq 0$, i.e., $\mu_p=0$ is not in the $\mu$-lattice, the eigenequation \eqref{eq:5} can be written as 
\begin{equation}\label{eq:eigenej1}
\begin{aligned}
 &\phi_{\omega_{p},\vec{k}}(\mathring \mu_p+\rho_p)=-\frac{\omega_{p}}{2\alpha}\phi_{\omega_{p},\vec{k}}(\mathring \mu_p),\\
&\phi_{\omega_{p},\vec{k}}(\mathring \mu_p-2\rho_p)=\frac{\omega_{p}}{2\alpha}\phi_{\omega_{p},\vec{k}}(\mathring \mu_p-\rho_p),\\
&\omega_{p}\,\phi_{\omega_{p},\vec{k}}(\mu_p)
+2\alpha\operatorname{sgn}(\mu_p)
\\
&\times\Big(\phi_{\omega_{p},\vec{k}}(\mu_p+\rho_p)
-\phi_{\omega_{p},\vec{k}}(\mu_p-\rho_p)
\Big)
\\
&=0,\ \forall \,|\mu_p|>\rho_p.
\end{aligned}
\end{equation}
\item[(ii)]  For the case of $\mathring \mu_p= 0$, i.e., $\mu_p=0$ is  in the $\mu$-lattice, one has
\begin{equation}\label{eq:eigenej2}
\begin{aligned}
 \omega_{p}\,\phi_{\omega_{p},\vec{k}}&(0)+\alpha\big(\phi_{\omega_{p},\vec{k}}(\rho_p)+\phi_{\omega_{p},\vec{k}}(-\rho_p)\big)=0,\\
&\omega_{p}\,\phi_{\omega_{p},\vec{k}}(\mu_p)
+2\alpha\operatorname{sgn}(\mu_p)
\\
&\times\Big(\phi_{\omega_{p},\vec{k}}(\mu_p+\rho_p)
-\phi_{\omega_{p},\vec{k}}(\mu_p-\rho_p)
\Big)
\\
&=0,\ \forall \, |\mu_p|>\rho_p.
\end{aligned}
\end{equation}
\end{itemize}

According to Eqs.~\eqref{eq:eigenej1} and \eqref{eq:eigenej2}, for $|\mu_p|>\rho_p$, the eigenequation becomes a constant--coefficient second-order difference equation which is solvable. With the convention
\begin{equation}
\beta:=\frac{\omega_{p}}{\mathrm{i}\alpha_{\rm eff}},\quad \alpha_{\rm eff}:=2\alpha=\frac{1}{2\sqrt{2\sqrt{3}\pi\gamma}\,\mathrm{i}\,l_p},
\end{equation}
the solution can be discussed case by case as follows. 
\begin{itemize}
\item[(i)] \textbf{Bounded oscillatory regime of \(|\beta|<2\).}
In this case, one can let $\beta=2\sin\theta$ with $\theta\in(-\tfrac{\pi}{2},\tfrac{\pi}{2})$. Then the solution reads
\begin{equation}
\label{eq:eigensol1}
\begin{aligned}
\phi_{\omega_{p},\vec{k}}(\mu_p)
&= a_{+}\,
\exp\!\Big(\mathrm{i}\theta\,\frac{\mu_p-\mathring{\mu}_p}{\rho_p}\Big)
\\
&\quad
+ a_{-}\,
\exp\!\Big(-\mathrm{i}\theta\,\frac{\mu_p-\mathring{\mu}_p}{\rho_p}\Big).
\end{aligned}
\end{equation}
where $a_\pm$ are constants.

\item[(ii)] \textbf{Threshold case of \(|\beta|=2\).} In this case, the solution reads 
\begin{equation}\label{eq:eigensol2}
\phi_{\omega_{p},\vec{k}}(\mu_p)=\Big(a_{+}+a_{-}\,\frac{\mu_p-\mathring \mu_p}{\rho_p}\Big)\,(\mathrm{sgn}(\beta)\mathrm{i})^{\frac{\mu_p-\mathring \mu_p}{\rho_p}},
\end{equation}
which grows at most polynomially in $|\mu_p|$.

\item[(iii)] \textbf{Unbounded regime of \(|\beta|>2\).} In this case, the solution is 
\begin{equation}\label{eq:eigensol3}
\phi_{\omega_{p},\vec{k}}(\mu_p)=a_+(r_+)^{\frac{\mu_p-\mathring \mu_p}{\rho_p}}+a_-(r_-)^{\frac{\mu_p-\mathring \mu_p}{\rho_p}}
\end{equation}
with 
\begin{equation}
r_{\pm}=\frac{\mathrm{i}\,\beta\pm \mathrm{i} \sqrt{\beta^2-4}}{2}.
\end{equation}
Because of $|r_+r_-|=1$, $(r_+)^{\frac{\mu_p-\mathring \mu_p}{\rho_p}}$ and $(r_-)^{\frac{\mu_p-\mathring \mu_p}{\rho_p}}$ exhibit the exponential growth in opposite directions: one diverges as $\frac{\mu_p-\mathring \mu_p}{\rho_p}\to \infty$ and the other diverges as $\frac{\mu_p-\mathring \mu_p}{\rho_p}\to -\infty$. Thus, $\phi(\mu_p)$ grows exponentially.
\end{itemize}

Note that a state in $\mathscr H^{g_p}_{\vec k}$ takes the form $\sum_{\vec\mu}\phi(\vec\mu)\,|g',\vec{k},\vec{\mu}\rangle$, and its normalizability in $\mathscr{H}^{g_p}_{\rm kin}$ requires $\sum_{\mu_p}|\phi(\mu_p)|^2$ to be finite. Thus, these solutions $\phi_{\omega_{p},\vec{k}}(\mu_p)$ defined by Eqs.~\eqref{eq:eigensol1}, \eqref{eq:eigensol2} and \eqref{eq:eigensol3} are not states in the Hilbert space. Nevertheless, the solutions given in Eqs.~\eqref{eq:eigensol1} and \eqref{eq:eigensol2} can be regarded as generalized eigenstates, in the sense that they belong to the dual space $\mathcal S'$ of some dense subspace $\mathcal S\subset \mathscr H^{g_p}_{\vec k}$. Here $\mathcal S$ can be chosen as the space of rapidly decaying sequences on the $\mu$-lattice, so that its dual $\mathcal S'$ contains oscillatory or at most polynomially growing solutions. This is valid in any super-selected subspace $\mathscr H^{g_p}_{\vec k}$ for the operator $\widehat\Theta_{\rm out}(p)$. In this sense, the states in $\mathscr{H}^{g_p}_{\rm kin}$ admit a spectral representation in terms of these generalized eigenstates.
In contrast, the solution given in Eq.~\eqref{eq:eigensol3} grows exponentially in $|\mu_p|$ and therefore does not define a continuous linear functional on $\mathcal{S}$, so it does not belong to $\mathcal{S}'$. Hence, those $\omega_{p}=\beta i\alpha_{\rm eff}$ with $|\beta|>2$ are not in the spectrum of $\widehat\Theta_{\rm out}(p)$. Consequently, the spectrum of $\widehat\Theta_{\rm out}$, corresponding to the generalized eigenstates given by Eqs.~\eqref{eq:eigensol1} and \eqref{eq:eigensol2} is continuous and locates in the following interval,
\begin{equation}\label{eq:band}
\sigma(\widehat\Theta_{\rm out}(p)) = \big[-2\,(\mathrm{i}\alpha_{\rm eff}),\,2\,(\mathrm{i}\alpha_{\rm eff})\big].
\end{equation}
It should be noted that only the second term in Eq.~\eqref{eq:ThetaOut-op-def} contributes the action of $\widehat\Theta_{\rm out}(p)$. Therefore, the same second term in Eq.~\eqref{eq:ThetaIn-op-def} will contribute the same part of the spectrum for the operator $\widehat\Theta_{\rm in}$.

\subsection{The spectra in the sector of $\Delta\neq0$}

In the sector of $\Delta\neq 0$, it is difficult to solve the eigen equation \eqref{eq:eigenequationv} analytically due to its non-trivial $\mu_j$-dependent coefficients.
Nevertheless, Eq.~\eqref{eq:theta1} implies that the operator $\widehat\Theta_{\rm out}$ can be decomposed in an natural way by separating the $\Delta$-independent shift part from the $\Delta$-dependent diagonal part as
\begin{equation}\label{eq:Theta-Delta-split}
\widehat{\Theta}_{\rm out}(v_j)=\widehat{\Theta}'(v_j)+\widehat{\Theta}_{\Delta}(v_j),
\end{equation}
where the action of $\widehat{\Theta}'(v_j)$ collects the off-diagonal shift terms in \eqref{eq:theta1} (the part that survives when $\Delta=0$), and the action of $\widehat{\Theta}_{\Delta}(v_j)$ contributes the remaining diagonal term proportional to $\Delta$. By construction, $\widehat{\Theta}'(v_j)$ is exactly a pure shift--type second-order difference operator on the $\mu$-lattice. In particular, its action has the same constant--coefficient structure as that of $\widehat{\Theta}_{\rm out}$ in the sector of $\Delta=0$ discussed in section \ref{sec:spectrum1}. Hence, the operator $\widehat{\Theta}_{\rm out}(v_j)$ has the same continuous (essential) spectral band as Eq.~\eqref{eq:band}. The operator $\widehat{\Theta}_{\Delta}(v_j)$ acts by multiplication on the $\mu$-lattice with
coefficient
\[
V(\mu_j)
= \frac{\operatorname{sgn}({\bar s_j})\Bigl(|\mu_j+\rho_j|^{\tfrac12}-|\mu_j-\rho_j|^{\tfrac12}\Bigr)^{2}}
{8\sqrt{3\gamma}\pi\,l_p\,
\bigl(|{\bar s_j+1}|^{\tfrac12}-|{\bar s_j-1}|^{\tfrac12}\bigr)}\Delta.
\]
Since one has $\bigl(|\mu_j+\rho_j|^{1/2}-|\mu_j-\rho_j|^{1/2}\bigr)^{2}=\mathcal{O}(|\mu_j|^{-1})$ and hence $V(\mu_j)$ as
$|\mu_j|\to\infty$ on the lattice, the operator $\widehat{\Theta}_{\Delta}^{(g_{v_j},\vec k)}$ is a compact self-adjoint operator in
$\mathscr H^{g_{v_j}}_{\vec k}$. Then the Weyl's theorem (see e.g. Ref.\cite{reed1980methods}) implies
the invariance of the essential spectrum under the perturbation
$\widehat{\Theta}'^{(g_{v_j},\vec k)}\mapsto \widehat{\Theta}'^{(g_{v_j},\vec k)}+\widehat{\Theta}_{\Delta}^{(g_{v_j},\vec k)}$, i.e.,
\begin{equation}\label{eq:ess-sector-sk}
\sigma_{\rm ess}\!\left(\widehat{\Theta}_{\rm out}^{(g_{v_j},\vec k)}\right)
=
\sigma_{\rm ess}\!\left(\widehat{\Theta}'^{(g_{v_j},\vec k)}\right).
\end{equation}
Therefore, the spectrum of
$\widehat{\Theta}_{\rm out}^{(g_{v_j},\vec k)}$ can differ from the shift band \eqref{eq:band} only by additional
discrete eigenvalues of finite multiplicity, and the only possible accumulation points of such
eigenvalues lie in the essential spectrum \cite{reed1980methods}.
As discussed in section \ref{sec:quantum}, the expansion operator can be expressed as  the direct sum:
\(
\widehat{\Theta}_{\rm out}(v_j)=\bigoplus_{g_{v_j}}\bigoplus_{\vec k}\widehat{\Theta}_{\rm out}^{(g_{v_j},\vec k)}.
\)
Hence, its spectrum and essential spectrum are obtained by
taking the closure of the union over all sectors as
\begin{equation}
\label{eq:Closure-ess-band}
\begin{aligned}
\sigma\!\left(\widehat{\Theta}_{\rm out}(v_j)\right)
&=
\overline{\bigcup_{g_{v_j},\vec k}
\sigma\!\left(\widehat{\Theta}_{\rm out}^{(g_{v_j},\vec k)}\right)},\\
\sigma_{\rm ess}\!\left(\widehat{\Theta}_{\rm out}(v_j)\right)
&=
\overline{\bigcup_{g_{v_j},\vec k}
\sigma_{\rm ess}\!\left(\widehat{\Theta}_{\rm out}^{(g_{v_j},\vec k)}\right)}.
\end{aligned}
\end{equation}
In particular, the essential spectrum is the shift band
\begin{equation}\label{eq:ess-band}
\sigma_{\rm ess}\!\left(\widehat{\Theta}_{\rm out}\right)
=
\big[-2\,(\mathrm{i}\alpha_{\rm eff}),\,2\,(\mathrm{i}\alpha_{\rm eff})\big].
\end{equation}
The remaining spectral points are sector-dependent discrete eigenvalues (isolated with finite
multiplicity), and any accumulation of such eigenvalues can occur only in
$\sigma_{\rm ess}\!\left(\widehat{\Theta}_{\rm out}\right)$. It should be noted that the above discussion on the spectrum of $\widehat{\Theta}_{\rm out}(v_j)$ can be applied in parallel to the operator $\widehat{\Theta}_{\rm in}$. {The fact that there is a sign difference between the first terms in Eqs.~\eqref{eq:ThetaOut-op-def} and \eqref{eq:ThetaIn-op-def} may only lead to the fact that $\widehat\Theta_{\rm out}^{g_{v_j},\vec k}$ and $\widehat\Theta_{\rm in}^{g_{v_j},\vec k}$ have different discrete spectra for the fixed graph $g$ and labels $\vec k$. However, the spectra of $\widehat\Theta_{\rm out}$ and $\widehat\Theta_{\rm in}$—defined as the direct sum over all admissible $g$ and $\vec k$—coincide. Indeed, as discussed below Eq.~\eqref{eq:theta2}, for any $\widehat\Theta_{\rm out}^{g_{v_j},\vec k}$ with fixed $\vec k$, one can always choose another label assignment $\vec k'$ such that $\widehat\Theta_{\rm in}^{g_{v_j},\vec k'}$ takes exactly the same operator form as $\widehat\Theta_{\rm out}^{g_{v_j},\vec k}$.} 

To further analyze the spectral properties of the expansion operators, we apply the following numerical approach \cite{zhang2019bouncing,zhang2020loop}. For any trial eigenvalue $\omega$, Eq.~\eqref{eq:eigenequationv} can be regarded as an iterative relation for determining $\phi(\mu_j+\rho_j)$. However, only for specific values of $\omega$ does the iteration produce bounded functions that qualify as legitimate generalized eigenfunctions. In other words, if the resulting function diverges for generic initial values, then $\omega$ is not an eigenvalue. Therefore, determining the spectrum of the expansion operator amounts to finding those values of $\omega$ for which the resulting $\phi(\mu_j)$ remain bounded. It is observed that the difference equation can be interpreted as an intruncated-matrix, denoted by $\mathsf{E}$, acting on the infinite-dimensional vector $\{\phi(\mu_j)\}$. Thus, to efficiently search for the allowed values of $\omega$, we introduce a cutoff to approximate $\mathsf{E}$ by a truncated-matrix $\mathsf{E}_n$, which we then diagonalize numerically. The spectral properties of $\mathsf{E}$ can be inferred by examining the behavior of the eigenvalues as the matrix size $n$ increases.

\begin{figure*}
  \centering
  \includegraphics[width=0.49\textwidth]{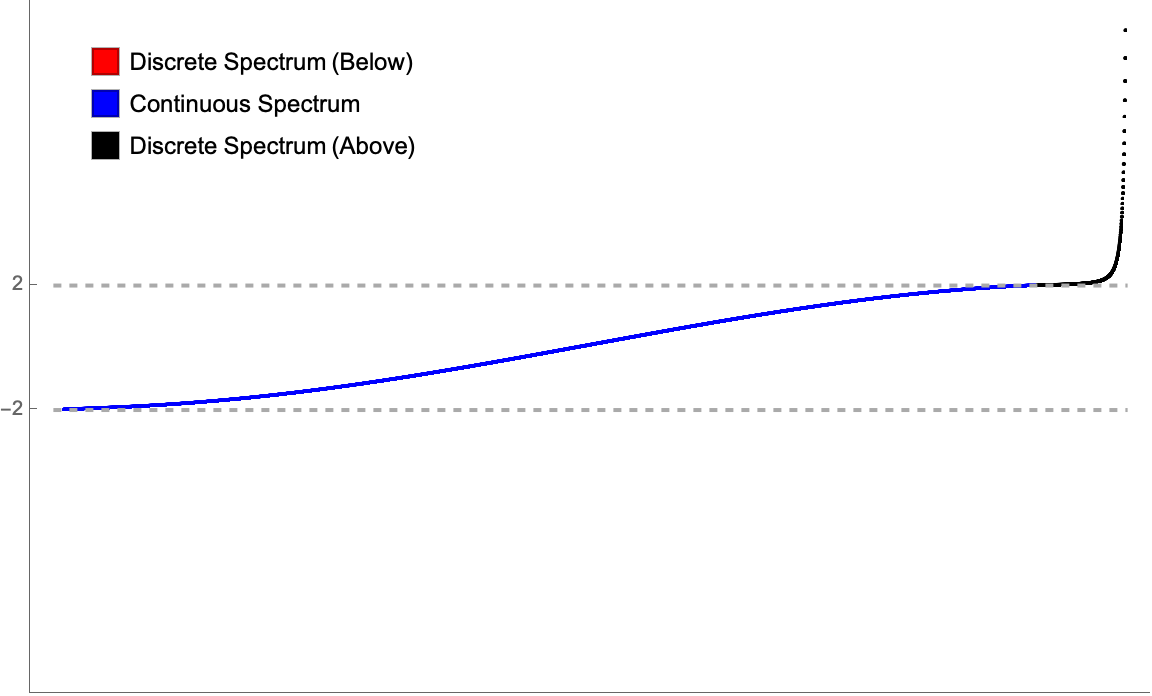}\hfill
  \includegraphics[width=0.49\textwidth]{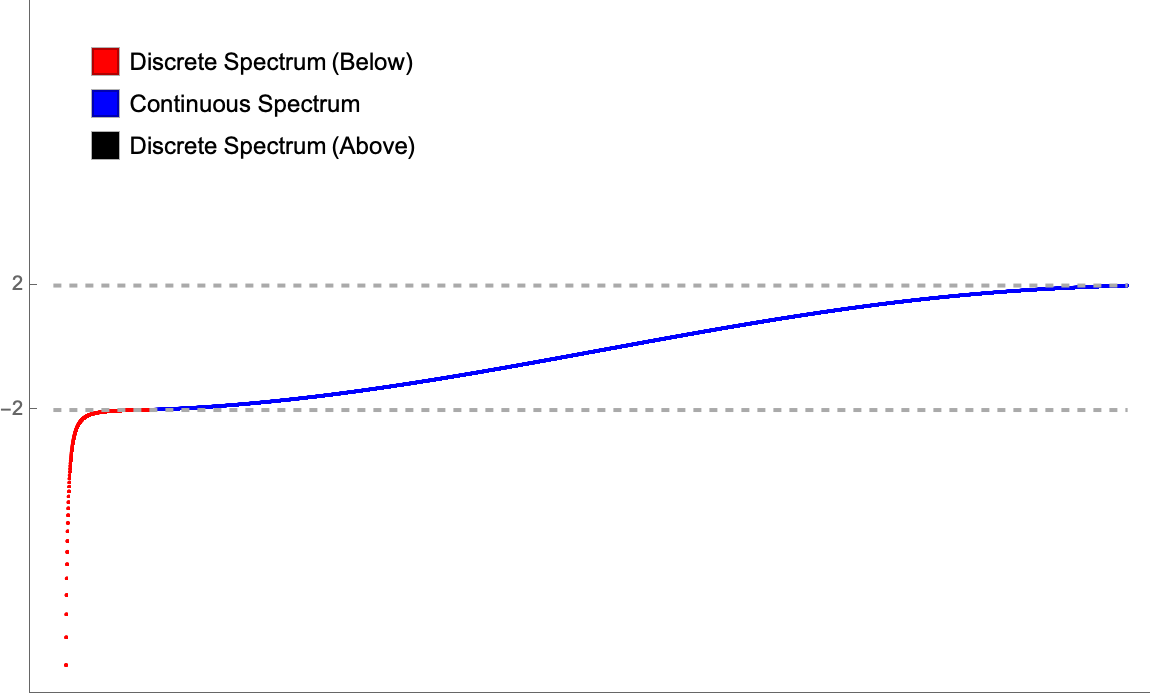}

  \caption{
  Numerical spectra of the expansion operators obtained from truncated matrix approximations ($n=10000$).
  (a) Spectrum of $\widehat{\Theta}_{\rm out}^{\,g_{v_j},\vec k}$.
  (b) Spectrum of $\widehat{\Theta}_{\rm in}^{\,g_{v_j},\vec k}$.
  The dashed lines indicate the analytic band edges at $\beta=\pm 2$.
  Discrete eigenvalues appear only above the band for $\widehat{\Theta}_{\rm out}^{\,g_{v_j},\vec k}$ and only below the band for
  $\widehat{\Theta}_{\rm in}^{\,g_{v_j},\vec k}$, while the continuous part fills the band interval.
  }
  \label{fig:expansion-spectra}
\end{figure*}

\begin{figure}[ht]
    \centering
    \includegraphics[width=0.47\textwidth]{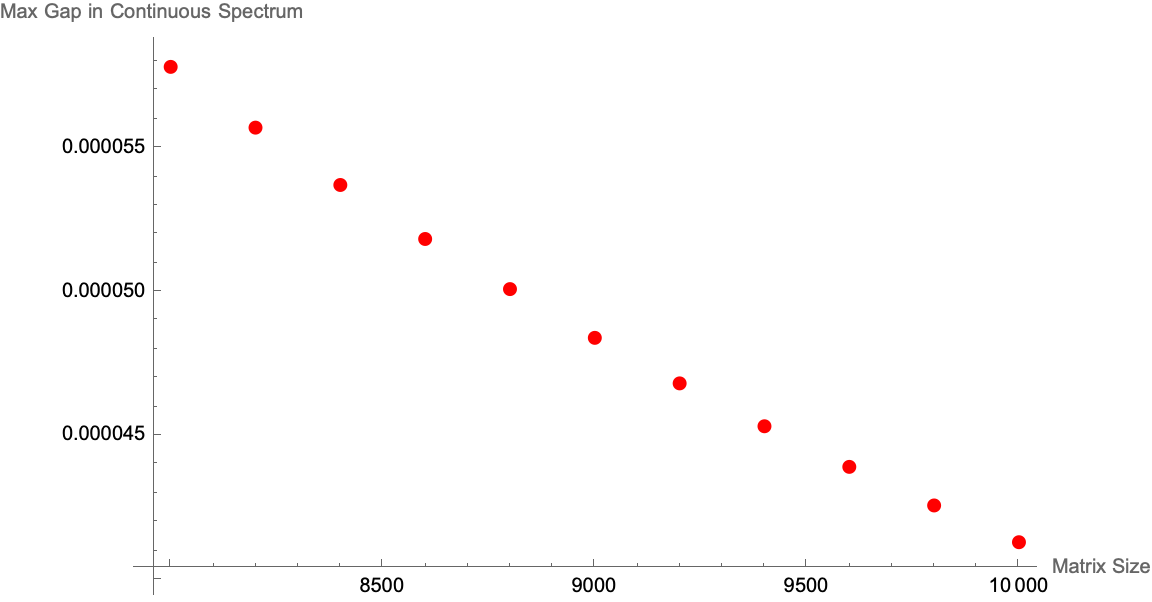}
    \caption{The maximum gap between adjacent eigenvalues of $\mathsf{E}$ in the interval $[-2,2]$ for different values of $n$.}
    \label{fig:eigen_spectrum_gap}
\end{figure}

\begin{figure}[ht]
    \centering
    \includegraphics[width=0.47\textwidth]{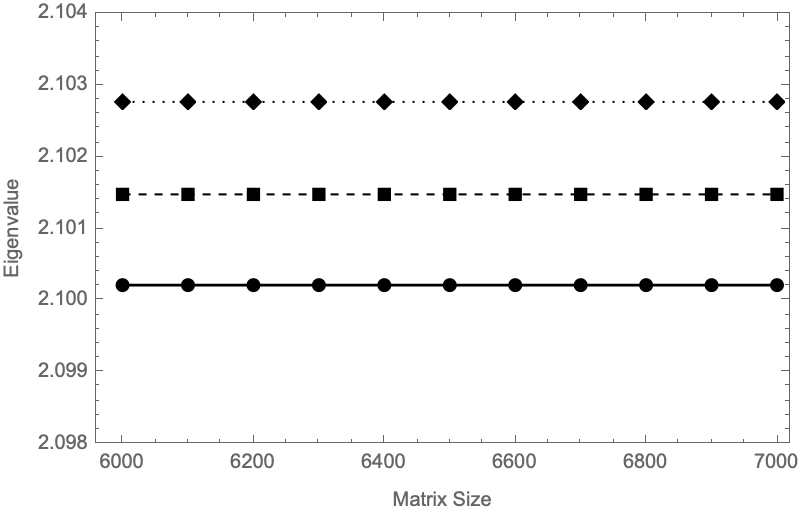}
    \caption{Convergence test for the truncated-matrix approximation of $\widehat{\Theta}_{\rm out}^{\,g_{v_j},\vec k}$: Three representative discrete eigenvalues lying above the continuous band are given as the truncation (matrix size) is increased. The observed plateaus indicate the numerical convergence of these discrete levels.}
    \label{fig:discrete_eigen_value_m}
\end{figure}

\begin{figure}
  \centering
  \includegraphics[width=0.9\linewidth]{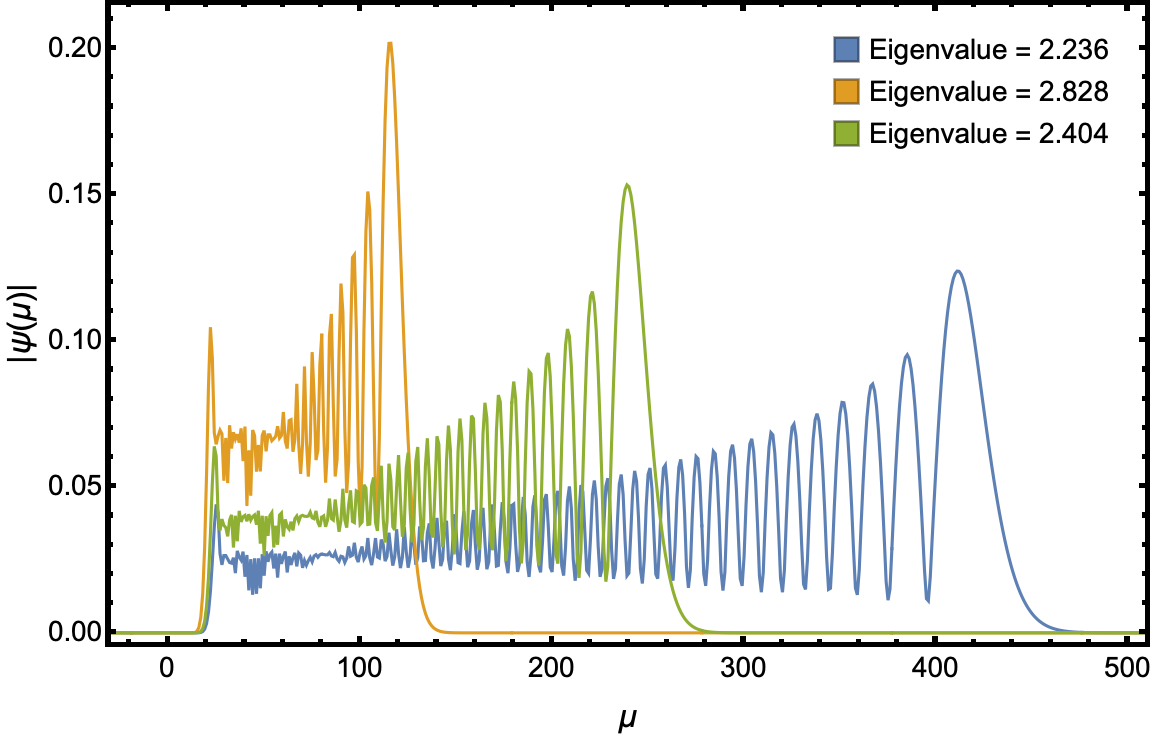}

  \vspace{1em}

  \includegraphics[width=0.9\linewidth]{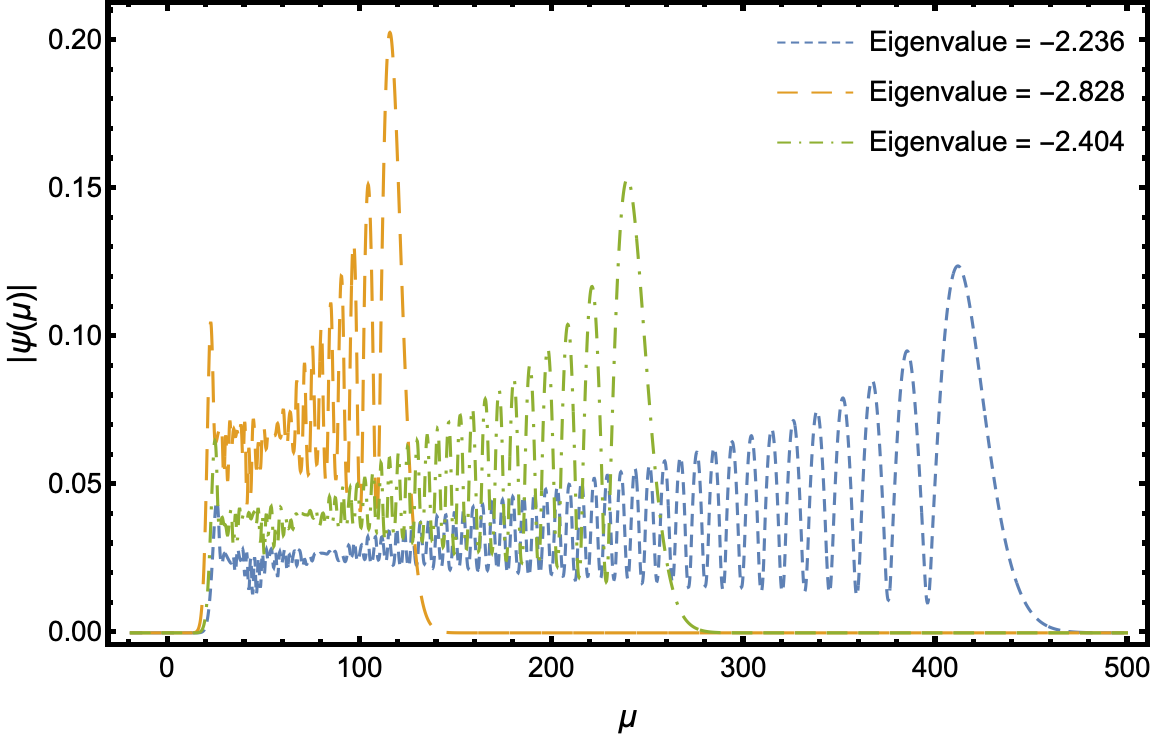}

  \caption{
  Representative localized eigenstates $\mathsf{E}_n$ associated with eigenvalues outside the band $[-2,2]$,
  obtained from truncated matrix approximations with $n=10000$.
  (a) Eigenstates of $\widehat{\Theta}_{\rm out}^{\,g_{v_j},\vec k}$ (top panel).
  (b) Eigenstates of $\widehat{\Theta}_{\rm in}^{\,g_{v_j},\vec k}$ (bottom panel).
  }
  \label{fig:eigenstates}
\end{figure}

To see why one can use the truncated-matrix $\mathsf{E}_n$ to approximate $\mathsf{E}$, let $\mu_j^{(0)}<0$ and $\mu_j^{(1)}>0$ be the left and right endpoints of the truncated $\mu$-lattice respectively.  Due to the imposed cutoff, the eigenequations for $\mathsf{E}_n$ differ from those of $\mathsf{E}$ at the boundaries $\mu_j^{(0)}$ and $\mu_j^{(1)}$. More, precisely, at the boundaries, Eq.~\eqref{eq:eigenequationv} implies that the eigenequations for $\mathsf{E}_n$ take the form 
\begin{equation}
\label{eq:eigen1}
\begin{aligned}
&(\mathsf{E}_n)_{\mu_j^{(0)},\mu_j^{(0)}}\,\phi(\mu_j^{(0)})
+(\mathsf{E}_n)_{\mu_j^{(0)},\mu_j^{(0)}+\rho_j}\,
\phi(\mu_j^{(0)}+\rho_j)
\\
&\qquad\qquad= \omega\,\phi(\mu_j^{(0)}),\\[0.4em]
&(\mathsf{E}_n)_{\mu_j^{(1)},\mu_j^{(1)}-\rho_j}\,
\phi(\mu_j^{(1)}-\rho_j)
+(\mathsf{E}_n)_{\mu_j^{(1)},\mu_j^{(1)}}\,
\phi(\mu_j^{(1)})
\\
&\qquad\qquad= \omega\,\phi(\mu_j^{(1)}).
\end{aligned}
\end{equation}
where we used the fact that $(\mathsf{E}_n)_{\mu_j,\mu_j'}=\mathsf{E}_{\mu_j,\mu_j'}$ and $\mathsf{E}_{\mu_j,\mu_j'}\neq 0$ only for $\mu_j=\mu_j',\mu_j'\pm \rho_j$, while the eigenequations for $\mathsf{E}$ read
\begin{equation}
\label{eq:eigen2}
\begin{aligned}
&\mathsf{E}_{\mu_j^{(0)},\mu_j^{(0)}-\rho_j}\,
\phi(\mu_j^{(0)}-\rho_j)
+\mathsf{E}_{\mu_j^{(0)},\mu_j^{(0)}}\,
\phi(\mu_j^{(0)})
\\
&+\mathsf{E}_{\mu_j^{(0)},\mu_j^{(0)}+\rho_j}\,
\phi(\mu_j^{(0)}+\rho_j)
= \omega\,\phi(\mu_j^{(0)}),\\[0.4em]
&\mathsf{E}_{\mu_j^{(1)},\mu_j^{(1)}-\rho_j}\,
\phi(\mu_j^{(1)}-\rho_j)
+\mathsf{E}_{\mu_j^{(1)},\mu_j^{(1)}}\,
\phi(\mu_j^{(1)})
\\
&+\mathsf{E}_{\mu_j^{(1)},\mu_j^{(1)}+\rho_j}\,
\phi(\mu_j^{(1)}+\rho_j)
= \omega\,\phi(\mu_j^{(1)}).
\end{aligned}
\end{equation}
Since the eigenstates of $E_n$ decay exponentially as $\mu_j$ approaches the boundaries $\mu_j^{(0)}$ and $\mu_j^{(1)}$ with sufficiently big absolute values, the additional terms appearing in Eq.~\eqref{eq:eigen2} but absent in Eq.~\eqref{eq:eigen1} can be neglected. Consequently, the truncated eigen equation \eqref{eq:eigen1} serves as a good approximation to the eigen equation \eqref{eq:eigen2}. Thus, the matrix size $n$ should be taken sufficiently large so that the eigenstates are sufficiently small at the boundaries $\mu_j^{(0)}$ and $\mu_j^{(1)}$, ensuring the validity of the approximation.

{In the following numerical study we work in a fixed-$\vec k$ sector and diagonalize the truncated matrix $\mathsf E_n$ associated with the corresponding difference equation.} As shown in Fig.~\ref{fig:expansion-spectra}, the eigenvalues obtained from the finite-dimensional approximation $\mathsf{E}_n$ {for both $\widehat{\Theta}_{\rm out}^{\,g_{v_j},\vec k}$ and $\widehat{\Theta}_{\rm in}^{\,g_{v_j},\vec k}$} can be categorized into two groups: the band of eigenvalues lies in $[-2,2]$ and the additional eigenvalues appear outside $[-2,2]$. Here the endpoints $\pm2$ arise from the chosen numerical normalization (the overall scale $\alpha_{\rm eff}$ is fixed in the computation), which has no bearing on the qualitative features of the spectrum. For the first group, as the cutoff $n$ increases, the level spacing inside the band decreases, as shown Fig.~\ref{fig:eigen_spectrum_gap}. This fact indicates that the corresponding spectrum of the operators is continuous and fills the interval $[-2,2]$. In contrast, for the eigenvalues in the second group, as illustrated in Fig.~\ref{fig:discrete_eigen_value_m}, they stabilize as $n$ increases and therefore correspond to discrete eigenvalues of the operator {$\widehat{\Theta}_{\rm out}^{\,g_{v_j},\vec k}$}. {The operator $\widehat{\Theta}_{\rm in}^{\,g_{v_j},\vec k}$ has similar discrete eigenvalues. In the examples shown in Fig.~\ref{fig:expansion-spectra}, these discrete eigenvalues appear on opposite sides of the band: on the upper side for $\widehat{\Theta}_{\rm out}^{\,g_{v_j},\vec k}$ and on the lower side for $\widehat{\Theta}_{\rm in}^{\,g_{v_j},\vec k}$.} Some eigenstates corresponding to the discrete eigenvalues of {$\widehat{\Theta}_{\rm out}^{\,g_{v_j},\vec k}$ and $\widehat{\Theta}_{\rm in}^{\,g_{v_j},\vec k}$} are computed separately. As illustrated in Fig.~\ref{fig:eigenstates}, the eigenstates of the two operators exhibit very similar envelope profiles. Moreover, the eigenstates decay exponentially when $|\mu_j|$ approaches $\infty$. As shown above, this feature is crucial in order to employ the eigenstates of $\mathsf{E}_n$ to approximate to the corresponding eigenstates of the full matrix $\mathsf{E}$.

\begin{figure}[ht]
    \centering
    \includegraphics[width=0.47\textwidth]{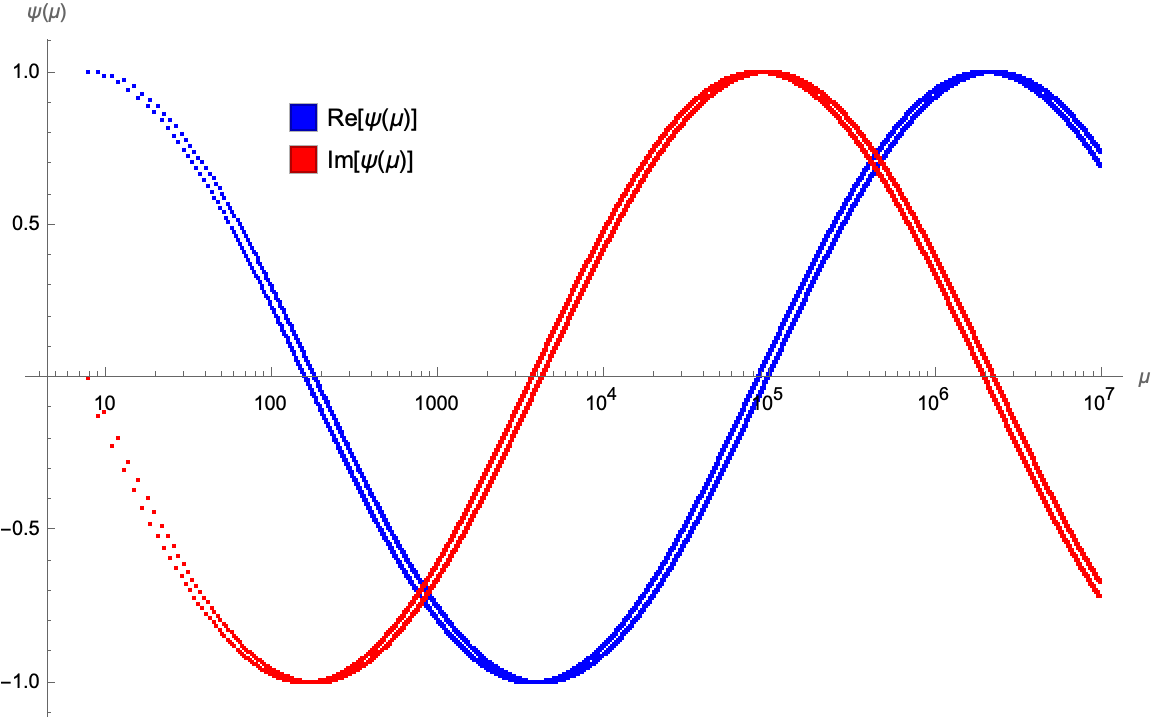}
    \caption{Eigenstate of $\widehat{\Theta}_{\rm out}^{\,g_{v_j},\vec k}(v_j)$ with eigenvalue $\omega=0$.}
    \label{fig:eigen state 0}
\end{figure}

In the classical theory, a marginally outer trapped surface is located at $\Theta_{\rm out}=0$. In this quantum model, it turns out that zero lies in the continuous spectrum of $\widehat{\Theta}_{\rm out}$, indicating that the associated eigenstate is non-normalizable. To determine this eigenstate more explicitly, we substitute $\omega_{v_j} = 0$ into Eq.~\eqref{eq:eigenequationv} and treat the resulting expression as a recurrence relation for $\phi(\mu_j+\rho_j)$. By numerical calculation, the corresponding eigenstate is shown in Fig.~\ref{fig:eigen state 0}, where its real and imaginary parts are plotted respectively as functions of $\mu$. It shows that the amplitude of the eigenstate remains bounded while exhibiting oscillatory behavior, similar to the plane wave eigenstates of the momentum operator in quantum mechanics.

\section{Summary and discussion}\label{sec:discussion}

To summarize, in previous sections the outward and inward null expansions of a two-sphere in the spatial manifold have been expressed in terms of the connection variables in the spherically symmetric geometry. This formulation enables us to quantize the null expansions in the spherically symmetric model of LQG. The actions of the outward expansion operator $\widehat{\Theta}_{\rm out}$ on the basis states in the kinematical Hilbert space $\mathscr H_{\rm kin}$ were given by Eqs.~\eqref{eq:theta1} and \eqref{eq:theta2}. By the orthonormal decomposition of $\mathscr H_{\rm kin}$ with respect to the graphs adapted to the expansion operator $\widehat{\Theta}_{\rm out}(v)$ at a point $v$, it has been shown that the operator is  self-adjoint in $\mathscr H_{\rm kin}$ and has generalized eigenstates. While our analysis is focused on the outward expansion operator $\widehat{\Theta}_{\rm out}$, it is applicable in parallel to the inward expansion operator $\widehat{\Theta}_{\rm in}$ due to the same constituents of the two operators. The spectra of the expansion operators have also been studied in details. Our analysis shows that the two operators have the same spectrum consisting of a continuous band and discrete points. In the subspaces $\mathscr H^{g_p}_{\vec k}$ with fixed labels $\vec{k}$, the two operators still share the common continuous spectrum, while their discrete spectra can be different from each other. In addition, for the restricted operators $\widehat\Theta_{\rm out}^{\,g_{v_j},\vec k}$ and $\widehat\Theta_{\rm in}^{\,g_{v_j},\vec k}$, one may find discrete eigenvalues lying outside the continuous band. In the examples shown in Fig.~\ref{fig:expansion-spectra}, these appear above (below) the band for $\widehat\Theta_{\rm out}^{\,g_{v_j},\vec k}$ $(\widehat\Theta_{\rm in}^{\,g_{v_j},\vec k})$, respectively. However, in the whole Hilbert space $\mathscr H_{\rm kin}$, the spectra of $\widehat\Theta_{\rm out}$ and $\widehat\Theta_{\rm in}$, as the union of the spectra in all subspaces $\mathscr H^{g_v}_{\vec k}$, coincide.

The results of this paper are enlightening. As shown in Sec.\ref{sec:quantum}, the restrictions of both $\widehat{\Theta}_{\rm out}(v)$ and $\widehat{\Theta}_{\rm in}(v)$ in $\mathscr H^{g_v}_{\vec k}$ are bounded and self-adjoint. Moreover, the two expansion operators are bounded at the points $p$ with $\Delta=0$ without restrictions. Note that all the inner points of any edge in a given graph underlying the cylindrical functions in $\mathscr H_{\rm kin}$ satisfy $\Delta=0$. In this sense both $\widehat{\Theta}_{\rm out}$ and $\widehat{\Theta}_{\rm in}$ are bounded almost everywhere in  $\mathscr H_{\rm kin}$. Recall that in the classical geometry the Raychaudhuri equation ties the expansions of null geodesic congruences to the curvature, and the divergence of the ingoing expansion is a necessary condition for the validity of the singularity theorems\cite{wald2010general}. The boundedness of expansion operators on almost all quantum states in our model gives a strong hint that the singularities of classical GR may be avoided in the quantum theory of gravity. Moreover, recall that the criteria for a classical apparent horizon are given by $\Theta_{\rm out}=0$ and $\Theta_{\rm in}<0$. Now both the outgoing and ingoing expansions have been quantized as the self-adjoint operators in $\mathscr H_{\rm kin}$ and their spectrums contain zero as well as certain negative numbers. These results provide the possibility to define a quantum version of the apparent horizon and thus set up the notion of quantum black holes. We leave this possibility for future investigations.

It should be noted that the $\bar\mu$-scheme is employed in this paper to construct the expansion operators. There are certainly other possible choices in constructing these operators. Different schemes for the regularization of the classical expansions may in general lead to inequivalent quantum operators. Thus, it is desirable to carry out the different quantization schemes and compare their quantum and semiclassical results. This open issue is also left for future investigations.

\acknowledgments

This  work is supported by the National Natural Science Foundation of China (NSFC) with Grants Nos. 12275022, 12505055 and 12405062, and ``the Fundamental Research Funds for the Central Universities''. G. L. is supported by the Guangdong Basic and Applied Basic Research Foundation with Grant No. 2026A1515012166.

\appendix
\section{Self-adjointness of the Expansion Operator}\label{app:appa}

Let $\mathscr H^{g_v}_{\vec k}$ be the closed subspace in $\mathscr H^{g_v}_{\rm kin}$ spanned by
the basis states $|g,\vec k,\vec\mu\rangle$ with $\vec k$ fixed.
Consider two generic vectors in $\mathscr H^{g_v}_{\vec k}$,
\begin{equation}
\begin{aligned}
|\varphi\rangle
&= \sum_{\mu_j} a(\mu_j,k_j)\,|\mu_j,k_j\rangle,\\
|\psi\rangle
&= \sum_{\mu_j} d(\mu_j,k_j)\,|\mu_j,k_j\rangle .
\end{aligned}
\end{equation}
and use the discrete inner product on the $\mu$-lattice,
\begin{equation}
\langle \psi|\varphi\rangle=\sum_{\mu_j}\overline{d(\mu_j,k_j)}\,a(\mu_j,k_j).
\end{equation}
In what follows we consider only the case of $\rho_j\neq 0$. 
For the case of $\rho_j=0$ (equivalently ${\bar s_j}=0$), Eq.~\eqref{eq:theta1} implies that the action $\widehat\Theta_{\rm out}^{(g_v,\vec k)}$ on the states gives the trivial  result. For the nontrivial action of $\widehat\Theta_{\rm out}^{(g_v,\vec k)}$, the coefficient of $|\mu_j,k_j\rangle$ in
$\widehat\Theta_{\rm out}^{(g_v,\vec k)}|\varphi\rangle$ is a sum of the following three contributions,
\begin{equation}
\begin{aligned}
b_1(\mu_j)
&= a(\mu_j, k_j)\,
\frac{\operatorname{sgn}({\bar s_j})}{\sqrt{\gamma}\,l_p\,\rho_j^2}
\Big(|\mu_j + \rho_j|^{\tfrac12} - |\mu_j - \rho_j|^{\tfrac12}\Big)^{2}
\\
&\quad \times
\Big(|{\bar s_j+1}|^{\tfrac12} - |{\bar s_j-1}|^{\tfrac12}\Big)\,\Delta .
\end{aligned}
\end{equation}
\begin{equation}
\begin{aligned}
b_2(\mu_j)
&= -\,a(\mu_j-\rho_j, k_j)\,
\frac{1}{2\,\mathrm{i}\sqrt{\gamma}\,l_p\,\rho_j}
\\
&\times\Big(|{\bar s_j+1}|^{\tfrac12}- |{\bar s_j-1}|^{\tfrac12}\Big)
c_{+}(\mu_j-\rho_j).
\end{aligned}
\end{equation}
\begin{equation}
\begin{aligned}
b_3(\mu_j)
&= a(\mu_j+\rho_j, k_j)\,
\frac{1}{2\,\mathrm{i}\sqrt{\gamma}\,l_p\,\rho_j}
\\
&\times\Big(|{\bar s_j+1}|^{\tfrac12}- |{\bar s_j-1}|^{\tfrac12}\Big)
c_{-}(\mu_j+\rho_j).
\end{aligned}
\end{equation}
where
\begin{equation}
\begin{aligned}
c_{+}(\mu)
&:= \operatorname{sgn}(\mu)+\operatorname{sgn}(\mu+\rho_j),\\
c_{-}(\mu)
&:= \operatorname{sgn}(\mu)+\operatorname{sgn}(\mu-\rho_j).
\end{aligned}
\end{equation}
satisfying $|c_{\pm}(\mu)|\le 2$ for all $\mu$. With these notations one has
\begin{equation}
\langle \psi|\,\widehat\Theta_{\rm out}^{(g_v,\vec k)}\,|\varphi\rangle
=\sum_{\mu_j}\overline{d(\mu_j,k_j)}\,
\Big(b_1(\mu_j)+b_2(\mu_j)+b_3(\mu_j)\Big).
\end{equation}
For the diagonal term, all prefactors multiplying $a(\mu_j,k_j)$ in $b_1(\mu_j)$ are real
(and $\operatorname{sgn}({\bar s_j})=\pm1$). Hence the diagonal contribution is manifestly symmetric.
For the shift terms, we rewrite the sums by changes of summation variables.
For the $b_2$ term, by setting $\nu=\mu_j-\rho_j$ we obtain
\begin{align}
&\sum_{\mu_j}\overline{d(\mu_j,k_j)}\,b_2(\mu_j)
=
-\frac{1}{2\,\mathrm{i}\sqrt{\gamma}\,l_p\,\rho_j}
\nonumber
\\
&\times\Big(|{\bar s_j+1}|^{\tfrac12}- |{\bar s_j-1}|^{\tfrac12}\Big)
\nonumber
\\
&\times\sum_{\nu}\overline{d(\nu+\rho_j,k_j)}\,a(\nu,k_j)\,c_{+}(\nu).
\label{eq:symm-shift-1}
\end{align}
For the $b_3$ term, by setting $\nu=\mu_j+\rho_j$ we obtain
\begin{align}
&\sum_{\mu_j}\overline{d(\mu_j,k_j)}\,b_3(\mu_j)
=
\frac{1}{2\,\mathrm{i}\sqrt{\gamma}\,l_p\,\rho_j}
\nonumber
\\
&\times\Big(|{\bar s_j+1}|^{1/2}
- |{\bar s_j-1}|^{1/2}\Big)
\nonumber
\\
&\times\sum_{\nu}\overline{d(\nu-\rho_j,k_j)}\,a(\nu,k_j)\,c_{-}(\nu).
\label{eq:symm-shift-2}
\end{align}
Using the identity
\begin{equation}
c_{-}(\nu)=\operatorname{sgn}(\nu)+\operatorname{sgn}(\nu-\rho_j)
=c_{+}(\nu-\rho_j),
\label{eq:cpm-identity}
\end{equation}
one sees that the shift contributions \eqref{eq:symm-shift-1}--\eqref{eq:symm-shift-2}
agree precisely with the corresponding contributions in
$\langle \widehat{\Theta}(v_j)\psi|\varphi\rangle$ through the same index shifts.
Therefore,
\begin{equation}
\langle \psi|\,\widehat\Theta_{\rm out}^{(g_v,\vec k)}\,|\varphi\rangle
=
\langle \widehat\Theta_{\rm out}^{(g_v,\vec k)}\psi|\varphi\rangle,
\qquad \forall\,|\varphi\rangle,|\psi\rangle\in\mathscr H^{g_v}_{\vec{k}},
\end{equation}
and hence $\widehat\Theta_{\rm out}^{(g_v,\vec k)}$ is symmetric in $\mathscr H^{g_v}_{\vec k}$.

Next we estimate the operator norm. The squared norm reads
\begin{equation}
\begin{aligned}
\bigl\|\widehat{\Theta}_{\rm out}^{(g_v,\vec k)}\,|\varphi\rangle\bigr\|^2
&= \sum_{\mu_j} \bigl|b_1(\mu_j)+b_2(\mu_j)+b_3(\mu_j)\bigr|^2 \\
&\le 3 \sum_{\mu_j} \Bigl(|b_1(\mu_j)|^2+|b_2(\mu_j)|^2+|b_3(\mu_j)|^2\Bigr).
\end{aligned}
\end{equation}
where we used $|x+y+z|^2\le 3(|x|^2+|y|^2+|z|^2)$. By denoting
\[
M(\mu_j):=|\mu_j+\rho_j|^{1/2}-|\mu_j-\rho_j|^{1/2},
\]
one has
\begin{equation}
\begin{aligned}
|b_1(\mu_j)|^2
&= |a(\mu_j,k_j)|^2\,\frac{\Delta^2}{\gamma\,l_p^2}\,
\frac{M(\mu_j)^4}{\rho_j^4}
\\
&\quad \times
\Big(|{\bar s_j+1}|^{\tfrac12} - |{\bar s_j-1}|^{\tfrac12}\Big)^{2}.
\end{aligned}
\end{equation}
where we used $\operatorname{sgn}({\bar s_j})^2=1$. Using the uniform bound
\[
\frac{M(\mu_j)^4}{\rho_j^{2}}\le 4,
\qquad
\rho_j^2=8\sqrt{3}\,\pi\,
\Big(|{\bar s_j+1}|^{1/2} - |{\bar s_j-1}|^{1/2}\Big)^{2},
\]
we obtain
\begin{equation}
|b_1(\mu_j)|^2
\le |a(\mu_j,k_j)|^2\,\frac{\Delta^2}{2\sqrt{3}\,\pi\,\gamma\,l_p^2}.
\end{equation}
Similarly, using $|c_{\pm}(\mu)|\le 2$ one has
\begin{align}
&|b_2(\mu_j)|^2
\nonumber
\\
&\le |a(\mu_j-\rho_j,k_j)|^2\,
\frac{1}{\gamma\,l_p^2\,\rho_j^2}
\Big(|{\bar s_j+1}|^{\tfrac12} - |{\bar s_j-1}|^{\tfrac12}\Big)^{2},\\[0.4em]
&|b_3(\mu_j)|^2
\nonumber
\\
&\le |a(\mu_j+\rho_j,k_j)|^2\,
\frac{1}{\gamma\,l_p^2\,\rho_j^2}
\Big(|{\bar s_j+1}|^{\tfrac12} - |{\bar s_j-1}|^{\tfrac12}\Big)^{2}.
\end{align}
and hence
\begin{align}
|b_2(\mu_j)|^2+&|b_3(\mu_j)|^2
\le \frac{1}{\gamma\,l_p^2\,\rho_j^2}\,
\Big(|{\bar s_j+1}|^{1/2} - |{\bar s_j-1}|^{1/2}\Big)^{2}
\nonumber\\
&\times\Big(|a(\mu_j-\rho_j,k_j)|^2+|a(\mu_j+\rho_j,k_j)|^2\Big)\notag\\
=\frac{1}{\gamma\,l_p^2}\cdot &\frac{1}{8\sqrt{3}\,\pi}\Big(|a(\mu_j-\rho_j,k_j)|^2
\nonumber+|a(\mu_j+\rho_j,k_j)|^2\Big),
\end{align}
where we used the relation between $\rho_j$ and ${\bar s_j}$ in the last equality.
Summing over $\mu_j$ and using the shift invariance
$\sum_{\mu_j} |a(\mu_j\pm\rho_j,k_j)|^2=\sum_{\mu_j} |a(\mu_j,k_j)|^2$, we get
\begin{equation}
\sum_{\mu_j} \big(|b_2(\mu_j)|^2+|b_3(\mu_j)|^2\big)
\le \frac{1}{4\sqrt{3}\,\pi\,\gamma\,l_p^2}\sum_{\mu_j} |a(\mu_j,k_j)|^2.
\end{equation}
Combining the estimates, we obtain
\begin{align}
&\|\widehat{\Theta}_{\rm out}^{(g_v,\vec k)}\,|\varphi\rangle\|^2
\nonumber
\\
&\le 3\sum_{\mu_j}|b_1(\mu_j)|^2+3\sum_{\mu_j}\big(|b_2(\mu_j)|^2+|b_3(\mu_j)|^2\big)\notag\\
&\le \left(\frac{3\,\Delta^2}{2\,\sqrt{3}\,\pi}
+\frac{3}{4\sqrt{3}\,\pi}\right)\frac{1}{\gamma\,l_p^2}\,\|\varphi\|^2,
\end{align}
and hence
\begin{equation}
\|\widehat{\Theta}_{\rm out}^{(g_v,\vec k)}\,|\varphi\rangle\|
\le \frac{1}{\sqrt{\gamma}\,l_p}\sqrt{\frac{6\,\Delta^2+3}{4\sqrt{3}\,\pi}}\;\|\varphi\|.
\end{equation}
This proves that $\widehat{\Theta}_{\rm out}^{(g_v,\vec k)}$ is bounded in $\mathscr H^{g_v}_{\vec k}$.
Since $\widehat{\Theta}_{\rm out}^{(g_v,\vec k)}$ is also symmetric, it is self-adjoint in $\mathscr H^{g_v}_{\vec k}$.

\end{document}